\documentclass[myepj-spec]{mySvjour}

\usepackage{graphicx}
\usepackage{hyperref}
\usepackage{color}
\usepackage{xspace}
\usepackage{natbib}   
\usepackage{verbatim}
\usepackage{amsfonts}
\usepackage{amssymb}
\definecolor{darkblue1}{rgb}{0,0,.2}
\definecolor{darkblue}{rgb}{0,0,.2}
\definecolor{darkred}{rgb}{0.5,0,0}
\pagecolor{white} 
\color{black}     
\hypersetup{breaklinks=true, 
            colorlinks=true, 
            linkcolor=darkblue1, 
            menucolor=darkblue1, 
            urlcolor=darkblue1,
            citecolor=darkblue1,
            pdftitle={},
            pdfauthor={},
            pdfsubject={},
            pdfkeywords={},
            pdfproducer={}
}
\bibstyle{plain}
\newcommand{\qbar}  {\ensuremath{\overline q}\xspace}

\newcommand{\ubar}  {\ensuremath{\overline u}\xspace}

\mathchardef\Upsilon="7107
\def\Y#1S{\ensuremath{\Upsilon{(#1S)}}\xspace}

\newcommand{\NNNLO}{\ensuremath{{\rm N}^3{\rm LO}}\xspace}
\newcommand{\mtau}{\ensuremath{m_\tau}\xspace}
\newcommand{\mZ}{\ensuremath{M_Z}\xspace}
\newcommand{\as}{\ensuremath{\alpha_{\scriptscriptstyle S}}\xspace}
\newcommand{\asTau}{\ensuremath{\as(\mtau^2)}\xspace}
\newcommand{\asmu}{\ensuremath{\as(\mu)}\xspace}
\newcommand{\ass}{\ensuremath{\as(s)}\xspace}
\newcommand{\asZ}{\ensuremath{\as(\mZ^2)}\xspace}
\newcommand{\asTauV}{\ensuremath{\as^{(V)}(\mtau^2)}\xspace}
\newcommand{\asTauA}{\ensuremath{\as^{(A)}(\mtau^2)}\xspace}
\newcommand{\asZZ}{\ensuremath{\as^{(Z)}(\mZ^2)}\xspace}
\newcommand{\asZtau}{\ensuremath{\as^{(\Tau)}(\mZ^2)}\xspace}
\newcommand{\MSb}{\ensuremath{\overline{\mathrm{MS}}}\xspace}

\newcommand\Ree{\ensuremath{R_{e^+e^-}}\xspace}
\newcommand\Rtau{\ensuremath{R_\tau}\xspace}
\newcommand\RtauV{\ensuremath{R_{\tau,V}}\xspace}
\newcommand\RtauA{\ensuremath{R_{\tau,A}}\xspace}
\newcommand\RtauS{\ensuremath{R_{\tau,S}}\xspace}
\newcommand\RtauVA{\ensuremath{R_{\tau,V/A}}\xspace}
\newcommand\RtauVpA{\ensuremath{R_{\tau,V+A}}\xspace}
\newcommand\GG{\ensuremath{\langle a_s GG\rangle}\xspace}
\newcommand\Osix{\ensuremath{\langle{\cal O}_6\rangle}\xspace}
\newcommand\Oeight{\ensuremath{\langle{\cal O}_8\rangle}\xspace}
\renewcommand\l{\ell}

\newcommand{\Kbar    }{\kern 0.2em\overline{\kern -0.2em K}{}\xspace}
\newcommand{\Kb      }{\ensuremath{\Kbar}\xspace}
\newcommand{\Kz      }{\ensuremath{K^0}\xspace}
\newcommand{\Kzb     }{\ensuremath{\Kbar^0}\xspace}
\newcommand{\KzKzb   }{\ensuremath{\Kz \kern -0.16em \Kzb}\xspace}
\newcommand{\Kp      }{\ensuremath{K^+}\xspace}
\newcommand{\Km      }{\ensuremath{K^-}\xspace}

\newcommand{\KpKm    }{\ensuremath{\Kp \kern -0.16em \Km}\xspace}

\newcommand{\Kstar   }{\ensuremath{K^\star}\xspace}

\def\NP{{Nucl. Phys.}\xspace}
\def\PL{{Phys. Lett.}\xspace}
\def\PR{{Phys. Rev.}\xspace}

\def\PRL{{Phys. Rev. Lett.}\xspace}

\def\ZP{{Z. Phys.}\xspace}
\def\EPJ{{Eur. Phys. J.}\xspace}

\newcommand{\FOPTmaybepp}{\ensuremath{{\rm FOPT}^{(++)}}\xspace}
\newcommand{\FOPTp}{\ensuremath{{\rm FOPT}^{+}}\xspace}
\newcommand{\FOPTpp}{\ensuremath{{\rm FOPT}^{++}}\xspace}

\newcommand{\CIPTpp}{\ensuremath{{\rm CIPT}^{++}}\xspace}

\newcommand{\third}{third\xspace}
\newcommand{\fourth}{fourth\xspace}
\newcommand{\fifth}{fifth\xspace}
\newcommand{\sixth}{sixth\xspace}
\newcommand{\nth}{$n$-th\xspace}

\newcommand{\e}{\varepsilon}
\newcommand{\hm}{\hspace{-0.1cm}}

\newcommand{\ointl}{\oint\limits}
\newcommand{\intl}{\int\limits}
\newcommand{\Tau}{\ensuremath{\tau}\xspace}
\newcommand{\nut}{\ensuremath{\nu_\tau}\xspace}
\newcommand{\nub}{\ensuremath{\overline{\nu}}\xspace}
\newcommand{\nueb}{\ensuremath{\nub_e}\xspace}
\newcommand{\BR}{\ensuremath{{\cal B}}\xspace}

\newcommand{\pim}{\ensuremath{\pi^-}\xspace}

\newcommand{\ee}{\ensuremath{e^+e^-}\xspace}

\newcommand{\Sew}{\ensuremath{S_{\rm EW}}\xspace}

\newcommand{\tev}{\ensuremath{\mathrm{\,Te\kern -0.1em V}}\xspace}
\newcommand{\gev}{\ensuremath{\mathrm{\,Ge\kern -0.1em V}}\xspace}
\newcommand{\mev}{\ensuremath{\mathrm{\,Me\kern -0.1em V}}\xspace}
\newcommand{\kev}{\ensuremath{\mathrm{\,ke\kern -0.1em V}}\xspace}
\newcommand{\ev}{\ensuremath{\mathrm{\,e\kern -0.1em V}}\xspace}
\newcommand{\gevc}{\ensuremath{{\mathrm{\,Ge\kern -0.1em V\!/}c}}\xspace}
\newcommand{\mevc}{\ensuremath{{\mathrm{\,Me\kern -0.1em V\!/}c}}\xspace}
\newcommand{\gevcc}{\ensuremath{{\mathrm{\,Ge\kern -0.1em V\!/}c^2}}\xspace}
\newcommand{\mevcc}{\ensuremath{{\mathrm{\,Me\kern -0.1em V\!/}c^2}}\xspace}

\newcommand{\beq}{\begin{equation}}
\newcommand{\eeq}{\end{equation}}
\newcommand{\beqn}{\begin{eqnarray}}
\newcommand{\eeqn}{\end{eqnarray}}
\newcommand{\beqns}{\begin{eqnarray*}}
\newcommand{\eeqns}{\end{eqnarray*}}
\newcommand{\bitm}{\begin{itemize}}
\newcommand{\eitm}{\end{itemize}}
\newcommand\pskip{\\[1.6ex]\noindent}
\renewcommand{\Re}{\ensuremath{{\rm Re}}\xspace}
\renewcommand{\Im}{\ensuremath{{\rm Im}}\xspace}

\newcommand\ie{{\it i.e.}\xspace}

\newcommand\cf{{\em cf.}\xspace}
\newcommand{\ea}{{\em et al.}\xspace}
\newcommand{\taum}{\tau^-}

\def\@citex[#1]#2{\if@filesw\immediate\write\@auxout{\string\citation{#2}}\fi
  \@tempcnta\z@\@tempcntb\m@ne\def\@citea{}\@cite{\@for\@citeb:=#2\do
    {\@ifundefined
       {b@\@citeb}{\@citeo\@tempcntb\m@ne\@citea
        \def\@citea{,\penalty\@m\ }{\bf ?}\@warning
       {Citation `\@citeb' on page \thepage \space undefined}}%
    {\setbox\z@\hbox{\global\@tempcntc0\csname b@\@citeb\endcsname\relax}%
     \ifnum\@tempcntc=\z@ \@citeo\@tempcntb\m@ne
       \@citea\def\@citea{,\penalty\@m}
       \hbox{\csname b@\@citeb\endcsname}%
     \else
      \advance\@tempcntb\@ne
      \ifnum\@tempcntb=\@tempcntc
      \else\advance\@tempcntb\m@ne\@citeo
      \@tempcnta\@tempcntc\@tempcntb\@tempcntc\fi\fi}}\@citeo}{#1}}

\def\@citeo{\ifnum\@tempcnta>\@tempcntb\else\@citea
  \def\@citea{,\penalty\@m}%
  \ifnum\@tempcnta=\@tempcntb\the\@tempcnta\else
   {\advance\@tempcnta\@ne\ifnum\@tempcnta=\@tempcntb \else
\def\@citea{--}\fi
    \advance\@tempcnta\m@ne\the\@tempcnta\@citea\the\@tempcntb}\fi\fi}
\newenvironment{myquote}
               {\list{}{\leftmargin0cm\indent}%
                \item\relax}
               {\endlist}
\newcommand\allFontSize{\footnotesize}
\newcommand\detailsSize{\allFontSize}
\newenvironment{details}%
{\begin{myquote}\detailsSize}{\end{myquote}}

\begin{document}

\headnote{}

\title{\boldmath The Determination of $\as$ from \Tau Decays Revisited}

\author{M.~Davier\inst{1} \and 
        S.~Descotes-Genon\inst{2} \and 
        A.~H\"ocker\inst{3} \and 
        B.~Malaescu\inst{1} \and 
        Z.~Zhang\inst{1}}

\institute{Laboratoire de l'Acc{\'e}l{\'e}rateur Lin{\'e}aire,
          IN2P3/CNRS et Universit\'e Paris-Sud 11 (UMR 8607), F--91405, Orsay Cedex, France \and
          Laboratoire de Physique Th\'eorique, 
          CNRS et Universit\'e Paris-Sud 11 (UMR 8627), 
          F--91405, Orsay Cedex, France \and
          CERN, CH--1211, Geneva 23, Switzerland}

\abstract{
We revisit the determination of \asTau using a fit to inclusive \Tau 
hadronic spectral moments in light of (1) the recent calculation of the 
\fourth-order perturbative coefficient $K_4$ in the expansion of the Adler
function, (2) new precision measurements from BABAR of \ee annihilation 
cross sections, which decrease the uncertainty in the separation of vector 
and axial-vector spectral functions, and (3) improved results from BABAR 
and Belle on \Tau branching fractions involving kaons. We estimate 
that the \fourth-order perturbative prediction reduces the theoretical
uncertainty, introduced by the truncation of the series, by 20\% with 
respect to earlier determinations. We discuss to some detail the perturbative 
prediction of two different methods: fixed-order perturbation theory (FOPT) and 
contour-improved perturbative theory (CIPT). The corresponding theoretical 
uncertainties are studied at the \Tau and $Z$ mass scales. The CIPT method is 
found to be more stable with respect to the missing higher order contributions
and to renormalisation scale variations. It is also shown that FOPT suffers from 
convergence problems along the complex integration contour. Nonperturbative 
contributions extracted from the most inclusive fit are small, in agreement 
with earlier determinations. Systematic effects from quark-hadron duality violation
are estimated with simple models and found to be within the quoted systematic errors. 
The fit based on CIPT gives $\asTau=0.344\pm0.005\pm0.007$, where the first error 
is experimental and the second theoretical. After evolution to \mZ we obtain
$\asZ=0.1212\pm0.0005\pm0.0008\pm0.0005$, where the errors are respectively 
experimental, theoretical and due to the evolution. The result is in 
agreement with the corresponding \NNNLO value derived from essentially the $Z$ width 
in the global electroweak fit. The \asZ determination from \Tau decays is the 
most precise one to date. 
}

\maketitle
\begin{flushright}
\normalsize
CERN-OPEN-2008-006, LAL 08-12, LPT-ORSAY 08-18, arXiv:0803.0979 \\
\today
\end{flushright}

\section{Introduction}

The relatively large mass of the \Tau lepton, its leptonic nature and its decay 
through weak interaction promotes it to a particular status for probing
the Standard Model (see~\cite{rmp} for a detailed review, and references therein). 
In particular, spectral functions determined from the invariant mass distributions of 
hadronic \Tau decays are fundamental quantities describing the production of hadrons 
from the non-trivial vacuum of strong interactions. They embed similar information to 
the one determined from cross sections of \ee annihilation to hadrons: both kinds of 
spectral functions are especially useful at low energies where perturbative QCD fails 
to locally describe the data, and where the theoretical understanding of the strong 
interactions remains at a qualitative level. Due to these limitations on the 
theoretical side, spectral functions play a crucial role in calculations of hadronic 
vacuum polarisation contributions to observables such as the effective electromagnetic 
coupling at the $Z$ mass, and the muon anomalous magnetic moment.
\pskip
Inclusive hadronic quantities, obtained after integrating over
the spectral functions (or directly via the measurement of hadronic or leptonic
\Tau branching fractions), have been found to be dominated by perturbative 
contributions at energies above $\sim$1\gev. They can be exploited to precisely
determine the strong coupling constant at the \Tau-mass scale, 
\asTau~\cite{narisonpich:1988,braaten:1989,bnp,pichledib}. More recently, 
this determination was reassessed~\cite{rmp} in the light of the existing 
data on $\tau$ decays and $e^+ e^-$ annihilation.
\pskip
In the present paper, we update the determination of \asTau from hadronic 
\Tau decays, motivated by progress performed in two different areas: on the 
theoretical side, the perturbative expression of the relevant correlator has 
been computed up to \fourth order~\cite{chetkuehn}, and on the experimental 
side, new precision measurements from BABAR of \Tau branching fractions involving 
kaons~\cite{babar-KKpi} decrease the uncertainty in the separation of vector and 
axial-vector spectral functions. We utilise this opportunity to analyse several 
features of the theoretical frameworks commonly used to determine \asTau in more 
detail. This concerns the treatments of the perturbative series, the convergence 
of the expansions, and the impact of nonperturbative effects.
\pskip
In Sec.~\ref{sec:def} we describe recent experimental improvement on the measurements 
of $K\Kb\pi$ decays, the spectral functions and the \Tau branching fractions. This is 
followed in Sec.~\ref{sec:theory} by a summary of the various theoretical prescriptions 
used to extract \asTau from a fit to data, and a discussion of their advantages and 
shortcomings. We also analyse the role played by nonperturbative contributions in this 
determination. In Sec.~\ref{sec:spectralmoments} we exploit the normalisation and shape of 
the spectral functions to constrain the relevant nonperturbative contributions and to 
provide an improved determination of \asTau. 

\section{Tau Hadronic Spectral Functions} 
\label{sec:def}


For vector (axial-vector) hadronic \Tau decay channels ${V^-}\nut$ (${A^-}\nut$),
the nonstrange vector (axial-vector) spectral function $v_1$ ($a_1$, $a_0$), where 
the subscript refers to the spin $J$ of the hadronic system, is derived from the 
invariant mass-squared distribution $(1/N_{V/A})(d N_{V/A}/d s)$ for a given hadronic 
mass $\sqrt{s}$, divided by the appropriate kinematic factor, and normalised to the 
hadronic branching fraction
\beq
\label{eq:sf}
   v_1(s)/a_1(s) 
   =
           \frac{\mtau^2}{6\,|V_{ud}|^2\,\Sew}\,
              \frac{\BR_{V^-/A^-\nut}}
                   {\BR_e} 
              \frac{d N_{V/A}}{N_{V/A}\,ds}\,
              \left[ \left(1-\frac{s}{\mtau^2}\right)^{\!\!2}\,
                     \left(1+\frac{2s}{\mtau^2}\right)
              \right]^{-1}\hspace{-0.3cm}.
\eeq
For $a_0(s)$, the same expression holds if the term $(1+2s/\mtau^2)$ is removed.
Here $\Sew=1.0198\pm0.0006$ is a short-distance electroweak 
correction~\cite{marciano,dehz02}, $\BR_{V^-/A^-\nut}$ 
($\BR_e$) denotes the inclusive $\tau\to V^-/A^-(\gamma)\nut$ 
($\tau\to e^-\nueb\nut$) branching fraction (throughout this letter, final 
state photon radiation is accounted for in the \Tau branching fractions). 
We use universality in the leptonic weak charged currents and the measurements
of $\BR_e$, $\BR_\mu$ and the $\tau$ lifetime, to obtain the improved branching
fraction $\BR_e=\BR_e^{\rm uni}=(17.818 \pm 0.032)\%$~\cite{rmp}.
We also use $\mtau=(1776.90 \pm 0.20)\mev$~\cite{pdg} 
and $|V_{ud}|=0.97418\pm0.00019$~\cite{ckmfitter} (assuming CKM unitarity). 
Integration of the spectral function over the \Tau phase space leads to the 
inclusive \Tau hadronic width, expressed through the ratio
\beq
   R_{\tau,V/A} = \frac{\BR_{V^-/A^-\nut}} {\BR_e}\,.
\eeq
\pskip
By unitarity and analyticity the spectral functions are connected to 
the imaginary part of the two-point correlation function, $\Pi_{ij,U}^{\mu\nu}(q)$,
for time-like momenta-squared $q^2>0$,
\beq
\label{eq:correl_def}
\Pi_{ij,U}^{\mu\nu}(q) \equiv i\int d^4x\,e^{iqx}
	\langle 0|T(U_{ij}^\mu(x)U_{ij}^\nu(0)^\dag)|0\rangle 
	=
	\left(-g^{\mu\nu}q^2+q^\mu q^\nu\right)
	\,\Pi^{(1)}_{ij,U}(q^2)+q^\mu q^\nu\,\Pi^{(0)}_{ij,U}(q^2)\,,
\eeq
where $U=A,V$ denotes the nature of the relevant currents, either 
vector $(U_{ij}^\mu= V_{ij}^\mu=\qbar_j\gamma^\mu q_i)$
or axial-vector ($U_{ij}^\mu= A_{ij}^\mu=\qbar_j\gamma^\mu\gamma_5 q_i$) 
charged colour-singlet quark currents. By Lorentz decomposition, the correlation 
functions can be split into their $J=1$ and $J=0$ parts. 
\pskip
In the complex $s=q^2$ plane, the polarisation functions $\Pi_{ij,U}^{\mu\nu}(s)$ 
are expected to exhibit a very simple analytic structure, the only non-analytic 
features being along the real axis: a branch cut for all polarisation functions, 
and a pole at the pion (kaon) mass for $a_0$. The imaginary part of 
the polarisation functions on the branch cut is linked to the spectral functions 
defined in Eq.~(\ref{eq:sf}), for nonstrange (strange) quark currents
\beq
\label{eq:imv}
   \Im\Pi^{(1,0)}_{\ubar d(s),V/A}(s) = \frac{1}{2\pi}v_1/a_{1,0}(s)\:,
\eeq
which provide the basis for comparing a theoretical description of strong interaction 
with hadronic data.
\pskip
Experimentally, the total hadronic observable \Rtau,
\beq
\label{eq:rtausum}
   \Rtau = \RtauV + \RtauA + \RtauS\:,
\eeq
where $\RtauS$ denotes the hadronic width to final states with net strangeness,
is obtained from the measured leptonic branching ratios, 
\beq
\label{rtau-lept}
\Rtau = \frac {1-\BR_e-\BR_\mu}{\BR_e} = \frac {1}{\BR_e^{\rm uni}}-1.9726 = 3.640 \pm 0.010\:.
\eeq

\subsection{New Input to the Vector/Axial-Vector Separation}

The separation of vector and axial-vector components is straightforward in the case 
of hadronic final states with only pions using $G$-parity, provided that isospin 
symmetry holds. An even number of pions has $G=1$ corresponding to vector states, 
while an odd number of pions has $G=-1$, which tags axial-vector states. 
Modes with a $K \Kb$ pair are not in general eigenstates 
of $G$-parity and contribute to both $V$ and $A$ channels. 
While the decay to $K^- K^0$ is pure vector, additional information is 
required to separate the $K \Kb \pi$ and the rarer $K \Kb \pi \pi$ modes. 
For the latter channel an axial-vector fraction of $0.5 \pm 0.5$ is used~\cite{rmp}.
\pskip
Until recently, there was some confusion on this issue for the 
$K \Kb \pi$ modes:

\begin{enumerate}

\item 	In the ALEPH analysis of $\tau$ decay modes with 
      	kaons~\cite{ALEPH:1999}, an estimate of the vector contribution was 
      	obtained using the $e^+e^-$ annihilation data from DM1~\cite{dm1} 
         and DM2~\cite{dm2} in the  
      	$K \Kb \pi$ channel, extracted in the $I=1$ state. 
      	This contribution was found to be small, and, using the conserved vector 
         current (CVC), a branching fraction of
      	$\BR_{\rm CVC} (\tau \to \nut (K \Kb \pi)_V) = (0.26 \pm 0.39)\cdot10^{-3}$, 
         was found, corresponding to an axial fraction
      	of $f_{A,{\rm CVC}}(K \Kb \pi)= 0.94^{+0.06}_{-0.08}$.

\item 	The ALEPH CVC result was corroborated by a partial-wave and 
      	lineshape analysis of the $a_1$ resonance from $\tau$ decays in the 
      	$\nut \pi^- 2\pi^0$ mode performed by CLEO~\cite{cleoa1}. 
      	The effect of the $\Kstar K$ decay mode of the $a_1$ was seen through
         unitarity and a branching fraction of
      	$\BR(a_1 \to \Kstar K) = (3.3 \pm 0.5)\%$ was derived. With the known
      	$\taum \to \nut a_1^-$ branching fraction, this value more than saturates
      	the total branching fraction available for the $K \Kb \pi$ channel, yielding
      	an axial fraction of $f_{A,a_1}(K \Kb \pi) = 1.30 \pm 0.24$.

\item 	Another piece of information, also contributed by 
         CLEO~\cite{CLEO:2004}, but conflicting with the two previous results, 
      	is based on a partial-wave analysis 
      	in the $K^- K^+ \pi^-$ channel using two-body resonance production and 
      	including many possible contributing channels. A much smaller axial 
      	fraction of $f_{A,K\Kb\pi}(K \Kb \pi) = 0.56 \pm 0.10$ was found here.

\end{enumerate}

\begin{figure}[t]
\centerline{\includegraphics[width=0.58\columnwidth]{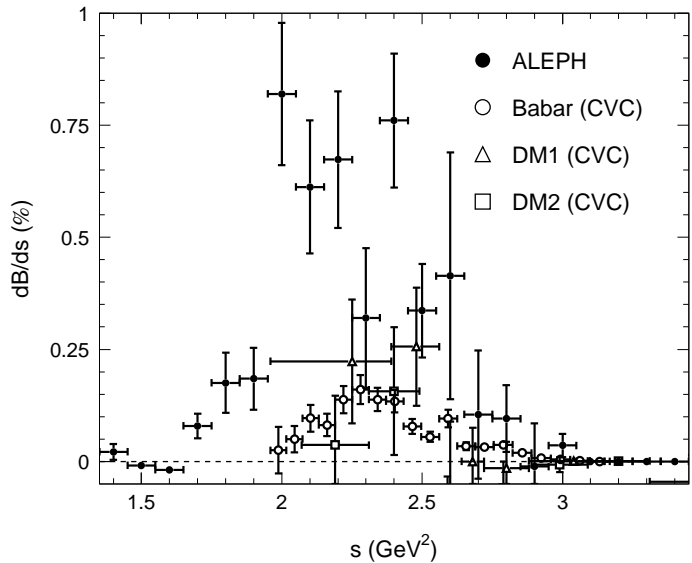}}
\vspace{-0.2cm}
\caption{\label{KKpi-CVC} The mass-squared distribution for $\tau\to \nut K \Kbar \pi$ decay 
         modes from ALEPH and the predictions for the vector component obtained by
         CVC using DM1, DM2 and BABAR $e^+e^-$ data.}
\end{figure}

Since the three determinations are inconsistent, the value $f_A = 0.75 \pm 0.25$ 
has been used previously to account for the discrepancy~\cite{rmp}. This led to 
a systematic uncertainty in the $V,A$ spectral functions that competed with 
the purely experimental uncertainties. 
\pskip
Precise cross section measurements for \ee annihilation to $K^+K^-\pi^0$
and to $K^0 K^\pm \pi^\mp$ have been recently published by the BABAR
Collaboration~\cite{babar-KKpi}, using the method of radiative return. In the mass 
range of interest for $\tau$ physics they show strong dominance of $K^\star(890)K$  
dynamics and a fit of the Dalitz plot yields a clean separation of the $I=0,1$
contributions. Assuming CVC, the mass distribution of the
vector final state in the decays $\tau \to \nut K \Kb \pi$ can be obtained. 
The result is shown in Fig.~\ref{KKpi-CVC} and compared with the full $\tau$ 
spectrum from ALEPH~\cite{ALEPH:1999} summing up the contributions from the
$K^- K^+ \pi^-$, $\Kzb\Kz \pi^-$, and $K^- \Kz \pi^0$
modes. The BABAR results reveal a small vector component. After integration, 
one obtains
\beq
\label{eq:Akkpi}
  f_{A,{\rm CVC}}(K \Kb \pi) = 0.833 \pm 0.024,
\eeq
which is about $1.3\sigma$ lower than the ALEPH determination using the same 
method (but with much poorer $e^+e^-$ input data) and 2.7$\sigma$ higher than 
the CLEO partial-wave-analysis result. The new determination has a precision 
that exceeds the previously used value by an order of magnitude, thus effectively
reducing the uncertainties in the vector and axial-vector spectral functions to the 
experimental errors only.
\pskip
One notices from Fig.~\ref{KKpi-CVC} that the axial fraction varies versus the
$K \Kb \pi$ mass, with lower masses being further axial-enhanced. The observed
axial-vector dominance is at variance with several estimates such as 
$f_A \sim 0.10$~\cite{gomez}, 0.37~\cite{roig}, obtained within
the Resonance Chiral Theory, which attempts at incorporating massive 
vector and axial resonances decaying into light mesons into a framework inspired by
chiral and large-$N_c$ arguments. On the other hand, this axial-vector dominance 
is closer to the prediction $f_A \sim 0.71$, based on a model combining axial-vector 
and vector resonances of finite widths with a leading-order chiral Lagrangian~\cite{finke}.
\pskip
In deriving Eq.~(\ref{eq:Akkpi}) care was taken to include a small contribution
from the $\phi \pi$ final state,  observed by BABAR in the same 
analysis~\cite{babar-KKpi}. Since BABAR also published a $\taum\to\nut\phi\pim$ 
branching fraction measurement~\cite{babar-tau-phipi}, it is possible to perform
a test of CVC in this channel with
\beqn
  \BR_{\rm CVC}(\tau \to \nut \phi \pi^-) &=& (3.8 \pm 0.9 \pm 0.2)\cdot10^{-5}\,, \\
  \BR_{\tau}(\tau \to \nut \phi \pi^-)    &=& (3.42 \pm 0.55 \pm 0.25)\cdot10^{-5}\,,
\eeqn
for which we find agreement within the quoted statistical and systematic errors.
For comparison the dominant CVC $\tau \to \nut K^\star(890)K$ 
branching fraction is $(7.3 \pm 0.6 \pm 0.4)\cdot10^{-4}$.
\begin{figure*}[t]
\centerline{
\includegraphics[width=0.49\columnwidth]{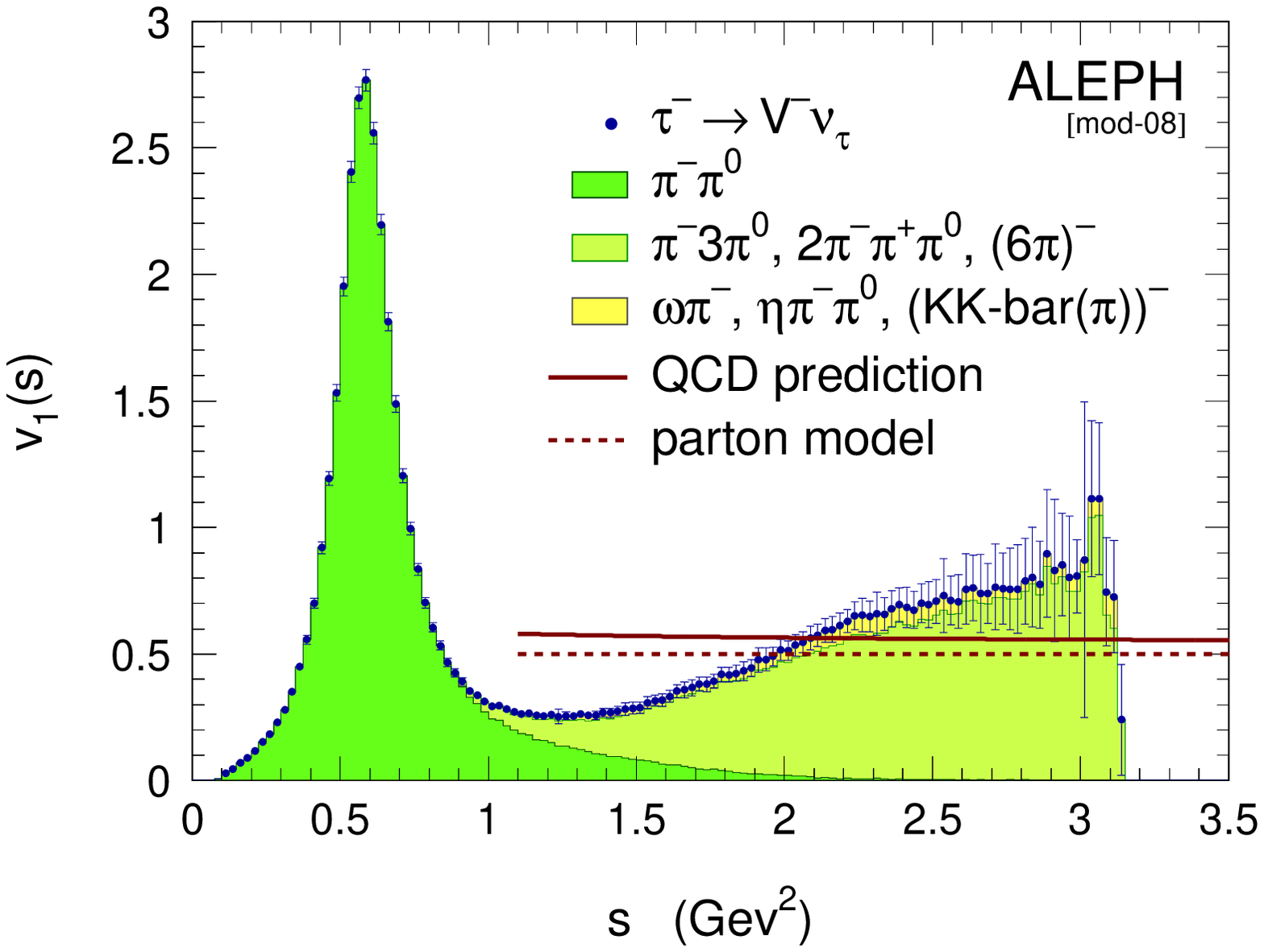}\hspace{0.2cm}
\includegraphics[width=0.49\columnwidth]{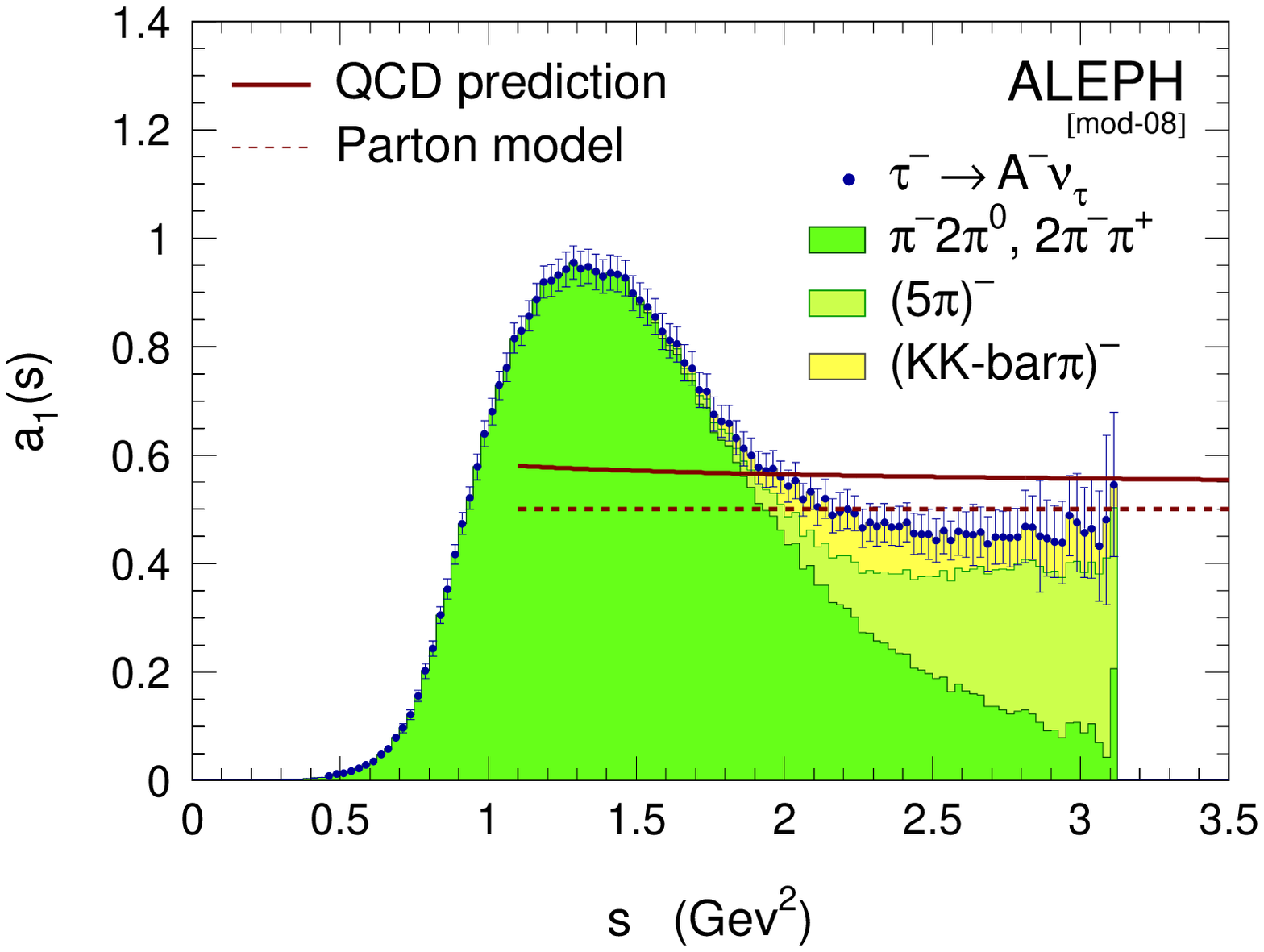}}
\vspace{-0.2cm}
\centerline{
\includegraphics[width=0.49\columnwidth]{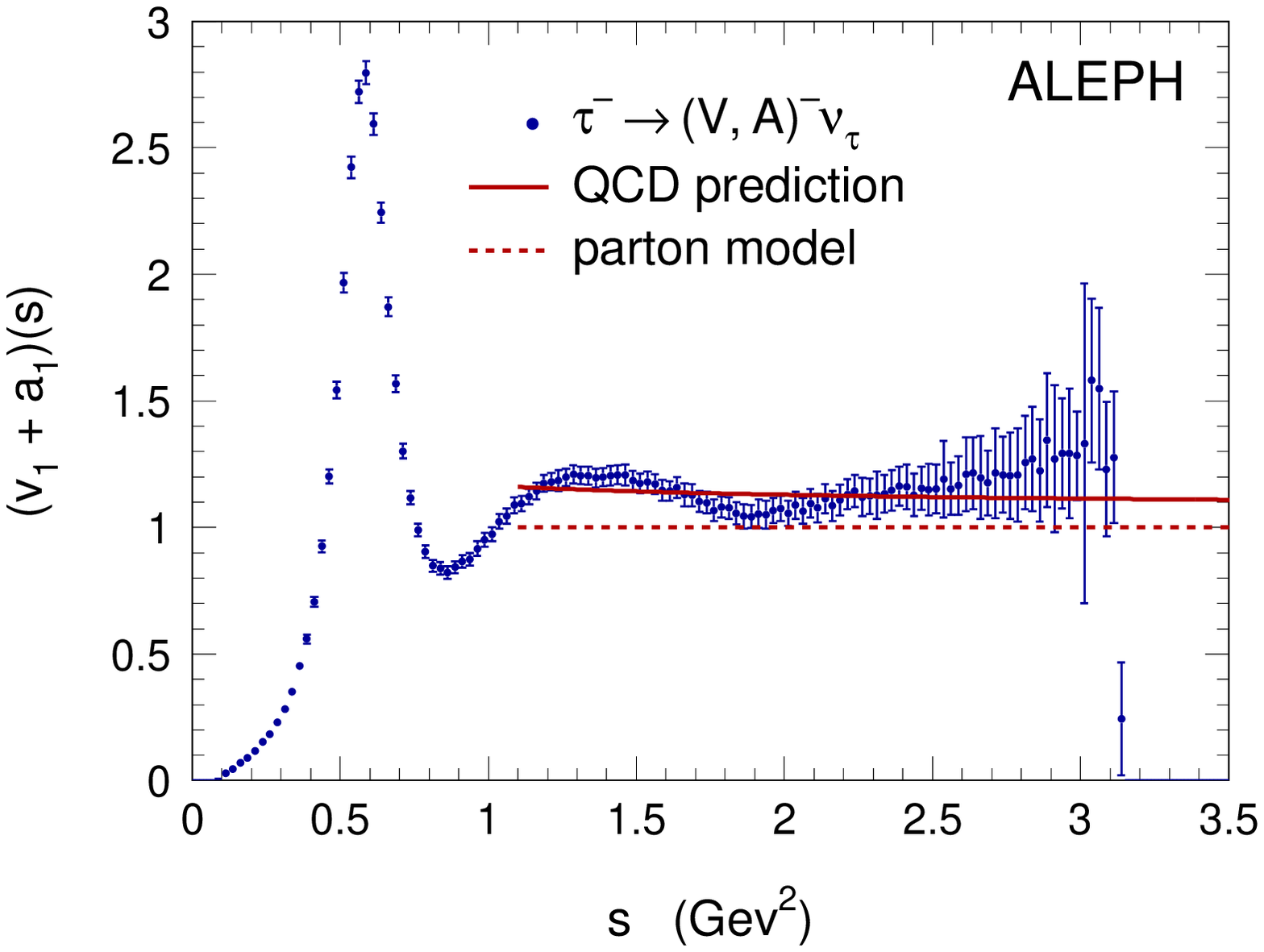}\hspace{0.2cm}
\includegraphics[width=0.49\columnwidth]{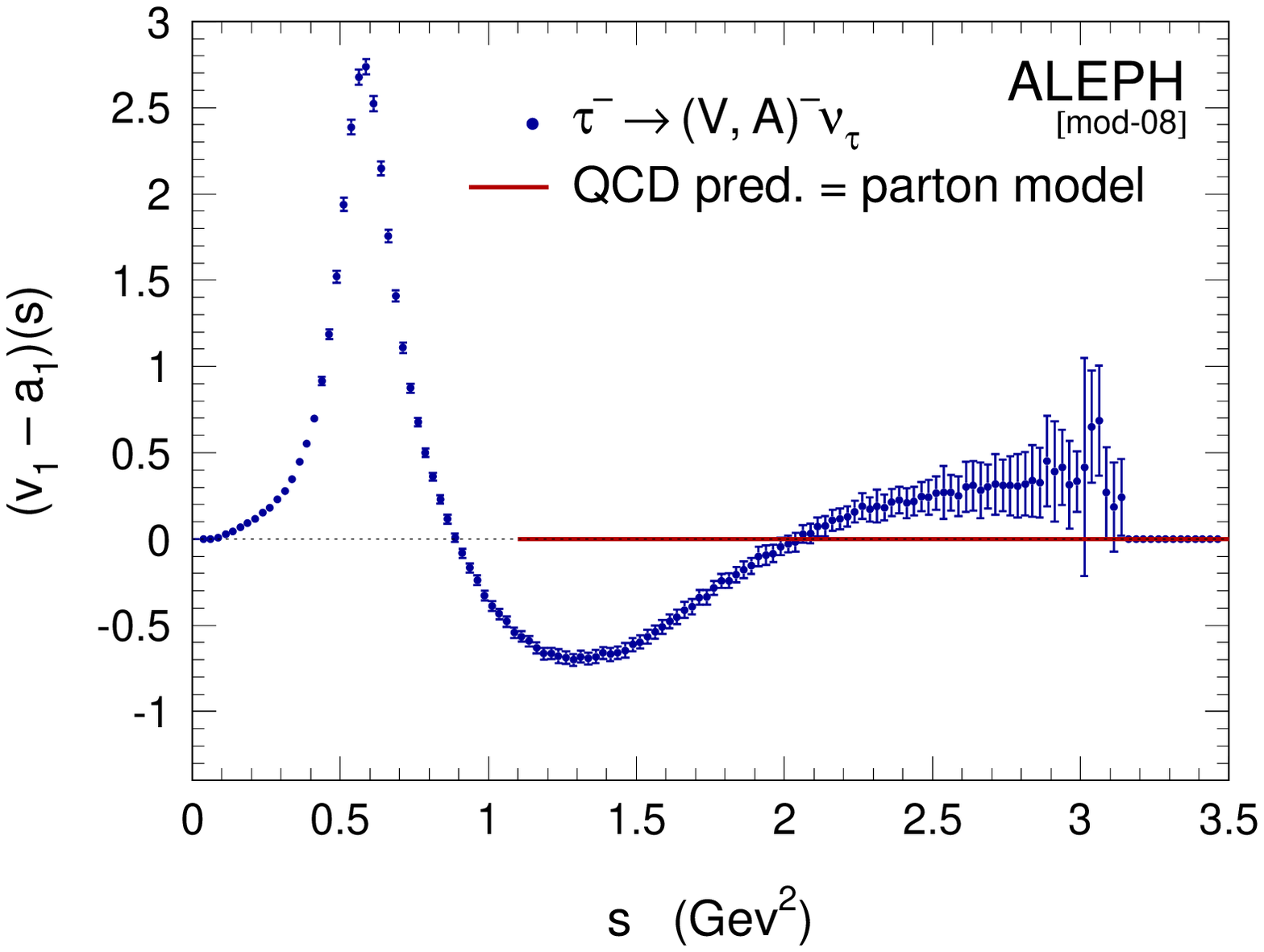}}
\vspace{-0.2cm}
\caption{Vector ($V$), axial-vector ($A$), $V+A$ and $V-A$ \Tau hadronic spectral 
         functions measured by ALEPH, and updated using the new $V,A$ separation
         in the $K \Kb \pi$ channels discussed in the text. 
         The shaded areas indicate the main contributing exclusive 
         $\tau$ decay channels. The curves show the predictions from the parton model 
         (dotted) and from massless perturbative QCD using $\asZ=0.120$ (solid).}
\label{fig:tauSF}
\end{figure*}

\subsection{Update on the Branching Fraction for Strange Decays}

New measurements of $\tau$ strange decays have been published since our last
compilation~\cite{rmp}. This is the case for the hadronic channels 
$K\pi^0$~\cite{babar-kpi0}, $K_S\pi^-$~\cite{belle-k0pi}, and
$K^-\pi^+\pi^-$~\cite{babar-KKpi}. Also using the more precise estimate
from universality for the $K^-$ channel~\cite{rmp}, the updated value
of \RtauS becomes
\beq
\label{rtau-s}
\RtauS = 0.1615 \pm 0.0040\:,
\eeq
replacing the previous value of $0.1666 \pm 0.0048$~\cite{rmp}. 
\pskip
Using the new $f_A(K \Kb \pi)$ value~(\ref{eq:Akkpi}), the updated hadronic 
widths $R_{\tau,V/A}$ from ALEPH, slightly renormalised so that their
sum agrees with the new average for $R_{\tau,V+A}$ obtained from 
(\ref{rtau-lept}) and (\ref{rtau-s}) read
\beqn
\label{eq:rtauv}
  \RtauV	&=& 1.783 \pm 0.011 \pm 0.002\:,\\
\label{eq:rtaua}
  \RtauA	&=& 1.695 \pm 0.011 \pm 0.002\:, \\
\label{eq:rtauvpa}
    R_{\tau,V+A}  &=& 3.479 \pm 0.011\:, \\
\label{eq:rtauvma}
    R_{\tau,V-A}  &=& 0.087 \pm 0.018 \pm 0.003\:,
\eeqn
where the first errors are experimental and the second due to the $V/A$ 
separation, now dominated by the $K \Kb \pi \pi$ channel.
\pskip
The ALEPH spectral functions are updated accordingly and shown in 
Fig.~\ref{fig:tauSF} for respectively vector, axial-vector, $V+A$ and $V-A$.

\section{Theoretical Prediction of \Rtau}
\label{sec:theory}

Tests of QCD and the precise measurement of the strong coupling constant \as at the 
\Tau mass scale~\cite{narisonpich:1988,braaten:1989,bnp,pichledib}, 
carried out first by the ALEPH~\cite{aleph_as} and 
CLEO~\cite{cleo_as} collaborations, have triggered many theoretical 
developments. They concern primarily the perturbative expansion
for which different optimised rules have been suggested. Among these 
are contour-improved (resummed) fixed-order perturbation 
theory~\cite{pert,pivov}, effective charge and minimal 
sensitivity schemes~\cite{grunberg1,grunberg2,dhar1,dhar2,pms}, the
large-$\beta_0$ expansion~\cite{beneke,altarelli,neubert}, as well as
combinations of these approaches. Their main differences lie
in how they deal with the fact that the perturbative series is truncated 
at an order where the missing part is not expected to be small. While a review
and discussion of the various approaches can be found in~\cite{rmp}, we only
recall some of their salient features in the following.
\pskip
With the publication of the full vector and axial-vector spectral 
functions by ALEPH~\cite{aleph_vsf,aleph_asf} and OPAL~\cite{opal_vasf}
it became possible to directly study the nonperturbative properties of 
QCD through $V-A$ sum rules and through fits to
spectral moments computed from weighted integrals over the spectral 
functions (we refer again to the discussions in~\cite{rmp}). 
Inclusive observables like \Rtau can be accurately predicted in terms of \asTau 
using perturbative QCD, and including small nonperturbative contributions 
within the framework of the Operator Product Expansion (OPE)~\cite{svz}. 

\subsection{Operator Product Expansion}	
\label{sec:qcd}

According to Eq.~(\ref{eq:imv}), the absorptive (imaginary) parts of the vector 
and axial-vector two-point correlation functions 
$\Pi^{(J)}_{\ubar d,V/A}(s)$, with the spin $J$ of the hadronic 
system, are proportional to the $\tau$ hadronic spectral functions  with 
corresponding quantum numbers. The nonstrange ratio \RtauVpA
can be written as an integral of these spectral functions  over the 
invariant mass-squared $s$ of the final state hadrons~\cite{bnp}
\beq
\label{eq:rtauth1}
\RtauVpA(s_0) =
	12\pi \Sew |V_{ud}|^2\! \intl_0^{s_0}
		\frac{ds}{s_0}\left(1-\frac{s}{s_0}
                                    \right)^{\!\!2}
     \left[\left(1+2\frac{s}{s_0}\right){\rm Im}\Pi^{(1)}(s+i\e)
      \,+\,{\rm Im}\Pi^{(0)}(s+i\e)\right],\hspace{0.2cm}\mbox{}
\eeq
where $\Pi^{(J)}$ can be decomposed as $\Pi^{(J)}=\Pi_{ud,V}^{(J)}+\Pi_{ud,A}^{(J)}$. 
We work in the chiral limit\footnote
{Vector and axial-vector currents are conserved in the chiral limit, 
	so that $s\Pi_V^{(0)}=s\Pi_A^{(0)}=0$.		
} 
to study the perturbative contribution, so that the lower integration limit 
is zero because of the pion pole at zero mass.
The correlation function $\Pi^{(J)}$ is analytic in the complex $s$ plane 
everywhere except on the positive real axis where singularities exist.
Hence by Cauchy's theorem, the imaginary part of $\Pi^{(J)}$ is 
proportional to the discontinuity across the positive real axis, and 
the integral~(\ref{eq:rtauth1}) can be replaced by a contour integral
over $\Pi(s)$ running counter-clockwise around the circle from $s=s_0+i\e$ 
to $s=s_0-i\e$.
\pskip
The energy scale $s_0= \mtau^2$ is large enough that contributions 
from nonperturbative effects are expected to be subdominant and the use 
of the Operator Product Expansion is appropriate. The latter is expected to
yield relevant results in the deep Euclidean region where $s$ is large and 
negative, whereas the extension to other regions in the 
complex plane is questionable. Fortunately, in the case of $\Rtau$,
the kinematic factor $(1-s/s_0)^2$ suppresses 
the contribution from the region near the positive real axis where 
$\Pi^{(J)}(s)$ has a branch cut and the validity of the OPE is doubtful
due to large quark-hadron duality violations~\cite{quinnetal,braaten88}. 
\pskip
The OPE of the vector and axial-vector ratio \RtauVA can be written as
\beq
\label{eq:delta}
   \RtauVA =
     \frac{3}{2}\Sew|V_{ud}|^2\bigg(1 + \delta^{(0)} + 
     \delta^\prime_{\rm EW} + \delta^{(2,m_q)}_{ud,V/A}
     +\hm\hm
     \sum_{D=4,6,\dots}\hm\hm\hm\hm\delta_{ud,V/A}^{(D)}\bigg)\:,
\eeq
with the massless universal\footnote
{In the chiral limit of vanishing quark masses the contributions from 
   vector and axial-vector currents coincide to any given order of 
   perturbation theory and the results are flavour independent.
} 
perturbative contribution $\delta^{(0)}$, the residual non-logarithmic 
electroweak correction $\delta^\prime_{\rm EW}=0.0010$~\cite{braaten} (\cf 
the discussion on radiative corrections in~\cite{rmp}), and the 
dimension $D=2$ perturbative contribution $\delta^{(2,m_q)}_{ud,V/A}$ 
from massive quarks. The term $\delta^{(D)}$ denotes the OPE contributions 
of mass dimension $D$~\cite{pichledib}
\beq
\label{eq:ope}
   \delta_{ud,V/A}^{(D)} =
      \hm\hm\hm\sum_{{\rm dim}{\cal O}=D}\hm\hm\hm C^\prime_{V/A}(s_0,\mu)
           \frac{\langle{\cal O}_{D}(\mu)\rangle_{V/A}}
                {s_0^{D/2}}\:,
\eeq
where
$\delta_{ud,V+A}^{(D)}=\frac{1}{2}\big(\delta_{ud,V}^{(D)}+\delta_{ud,A}^{(D)}\big)$. 
In practice, the OPE provides a separation between short and long 
distances by following the flow of a large incoming momentum. The scale 
parameter $\mu$ separates the long-distance nonperturbative effects, 
absorbed into the vacuum expectation value of the operators $\langle{\cal 
O}_{D}(\mu)\rangle$, from the short-distance effects that are included in 
the coefficients $C_{V/A}(s,\mu)$, which become $C^\prime_{V/A}(s_0,\mu)$ after 
performing the integration~(\ref{eq:rtauth1}).
The vacuum expectation values $\langle{\cal O}_{D}(\mu)\rangle$ encode 
information on the nonperturbative features of QCD vacuum and its effects 
on the propagation of quarks: they cannot be computed from first 
principles and have to be extracted from data. The short-distance 
coefficients $C_{V/A}(s,\mu)$ can be determined within perturbative QCD.

\subsection{Perturbative Contribution to Fourth Order in \as}
\label{sec:pert}

\Rtau is a doubly inclusive observable since it is the result of an 
integration over all hadronic final states at a given invariant mass and further 
over all masses between $m_\pi$ and $\mtau$. The scale $\mtau$ lies in a compromise 
region where \asTau is large enough so that \Rtau is sensitive to its value, 
yet still small enough so that the perturbative expansion converges 
safely and nonperturbative power terms are small. The prediction 
for $\Rtau$ is thus found to be dominated by the lowest-dimension term in 
Eq.~(\ref{eq:ope}), \ie, the term obtained from a perturbative computation 
of the correlator $\Pi$.
\pskip
For the evaluation of the perturbative series, it is convenient to 
introduce the analytic Adler function~\cite{adler} 
$D(s) \equiv - s\cdot d\Pi(s)/ds$, which avoids extra subtractions
that are unrelated to QCD dynamics.
The function $D(s)$ calculated in perturbative QCD within the \MSb 
renormalisation scheme is a function of \as and depends on the renormalisation 
scale $\mu$, occurring through $\ln(\mu^2/s)$. Since $D(s)$ is connected to a 
physical quantity, the spectral function  $\Im\Pi(s)$, it cannot depend on the 
choice of the renormalisation scale $\mu$. This is achieved through the 
cancellation of the $\mu$-dependence of \as\ and of the explicit occurrences
of $\mu$ in $D$.
Nevertheless, in the realistic case of a series truncated at a given order
in \as\ our knowledge of the renormalisation scale dependence is imperfect,
\ie, $D$ depends on $\mu$, thus inducing a systematic uncertainty. 
\pskip
To introduce the Adler function in Eq.~(\ref{eq:rtauth1}),
one uses partial integration, giving
\beq
\label{eq:rtauadler}
   1+\delta^{(0)} = 
      -2\pi i\hm\hm\ointl_{|s|=s_0}\hm\hm\frac{ds}{s}w(s)D(s)\:,
\eeq
where $w(s)=1-2s/s_0 + 2(s/s_0)^3-(s/s_0)^4$.
The perturbative expansion of $D(s)$ reads
\beq
\label{eq:adlerpert}
   D(s) = \frac{1}{4\pi^2} \sum_{n=0}^\infty \tilde{K}_n(\xi)a_s^{n}(-\xi s)\:,
\eeq
with $a_s\equiv\as/\pi$, and where the dimensionless factor $\xi$ parametrises the 
renormalisation scale ambiguity. While the coefficients $K_{0,1}=\tilde{K}_{0,1}=1$ 
are universal (we use the notation $K_n=\tilde{K}_n(\xi=1)$ in the following), the 
$\tilde{K}_{n\geq2}$ depend on the renormalisation scheme and scale used.
Powerful computational techniques have recently allowed to determine $K_4$.
The authors of~\cite{chetkuehn} exploited the dependence of the four-loop master 
integrals (used to express all relevant four-loop integrals with massless 
propagators) on the space-time dimension to compute the integrals to the 
required accuracy. For $n_f=3$ quark flavours and $\xi=1$ one has\footnote
{The numerical expressions for an arbitrary number of quark flavours ($n_f$) in 
   the \MSb renormalisation scheme for $\xi=1$ are:
   $K_0 = 1$,
   $K_1 = 1$,
   $K_2 \simeq 1.9857 - 0.1153\,n_f$,
   $K_3 \simeq 18.2428 - 4.2158\,n_f + 0.0862\,n_f^2$, and
   $K_4 \simeq 135.7916 - 34.4402\,n_f + 1.8753\,n_f^2 - 0.0101\,n_f^3 $.
}
$K_2\simeq1.640$, $K_3\simeq6.371$ and 
$K_4\simeq49.08$~\cite{adler1,adler2,2loop,loopbis,loopbisbis,chetkuehn}.
The full expressions for the functions $\tilde{K}_n(\xi)$ for arbitrary $\xi$ up 
to order $n=5$ can be found in~\cite{rmp}.
\pskip
With the series~(\ref{eq:adlerpert}), inserted into the r.h.s. of Eq.~(\ref{eq:rtauadler}), 
one obtains the perturbative expansion
\beq 
\label{eq:knan}
   \delta^{(0)} = 
       \sum_{n=1}^\infty \tilde{K}_n(\xi) A^{(n)}(a_s)\:,
\eeq
with the functions~\cite{pert}
\beq
\label{eq:an}
   A^{(n)}(a_s) 
	=
      \frac{1}{2\pi i}\hm\ointl_{|s|=s_0}\hm\hm
      \frac{ds}{s}w(s)a_s^{n}(-\xi s)
	=
  	   \frac{1}{2\pi} \intl_{-\pi}^{\pi}
      d\varphi\, w(-s_0e^{i\varphi})a_s^{n}(\xi s_0e^{i\varphi})\,. 
\eeq
\pskip
Similarly, the Adler function also serves to obtain the perturbative expansion 
of the inclusive \ee annihilation cross section ratio 
\beq
   \Ree(s)=\frac{\sigma(\ee\to{\rm hadrons}\,(\gamma))}{\sigma(\ee\to\mu^+\mu^-)}
          =-6\pi i \sum_fQ_f^2 \oint_{|s^\prime|=|s|}ds^\prime\cdot \frac{D(s^\prime)}{s^\prime}. 
\eeq
Evaluating the contour integral in fixed-order perturbation theory (\cf 
Sec.~\ref{sec:FoptAndCipt}) with $n_f=5$ active quark flavours, and 
inserting all known coefficients, gives\footnote
{The explicit formula reads:
 \begin{eqnarray}\Ree(s)&=& 3 \sum_f Q_f^2 \bigg[ 1 + a_s(s) + K_2\, a_s^2(s)  +
   \left(K_3 - \frac{1}{3}\pi^2\beta_0^2\right) a_s^3(s)
   + \left(K_4 - \frac{5}{6}\pi^2\beta_0\beta_1 - K_2\pi^2\beta_0^2\right) a_s^4(s)\nonumber\\[-0.2cm] \nonumber
&&\quad\quad\quad\quad +
   \left( K_5- \frac{1}{2} \pi^2 \beta_1^2 - \pi^2 \beta_0\beta_2 - \frac{7}{3}\pi^2 \beta_0\beta_1 K_2 - 2 \pi^2 \beta_0^2 K_3 \right. +
   \left. \frac{1}{5} \pi^4 \beta_0^4 \right) a_s^5(s) + \dots \bigg]\,. \end{eqnarray} 
}
\beqn
\Ree^{(5)}(s)&=& 3{\sum}_f Q_f^2 \big[  1 + a_s(s) + 1.4092\, a_s^2(s)
              - 12.7673\, a_s^3(s) - 79.9795\, a_s^4(s)  \nonumber\\           
              && \hspace{1.5cm}
              +\; (K_5 + 79.7306)\, a_s^5(s) + (K_6 + 2202.78)\, a_s^6(s) + \dots\big]\,.
\eeqn

\subsubsection{Fixed-Order and Contour-Improved Perturbation Theory}
\label{sec:FoptAndCipt}

The standard perturbative method to compute the contour integral consists of expanding
all the quantities up to a given power of $a_s(s_0)$. The starting point is the 
solution of the renormalisation group equation (RGE) for $a_s(s)$, which is expanded 
in a Taylor series of $\eta\equiv\ln(s/s_0)$ around the reference scale $s_0$~\cite{rmp}
\beqn
\label{eq:astaylor}
a_s(s) &=&
  		a_s - 
		\beta_0 \eta a_s^2 + 
		\left(-\beta_1\eta + \beta_0^2 \eta^2\right) a_s^3 +
    	\bigg(-\beta_2\eta  
        + \frac{5}{2} \beta_0\beta_1 \eta^2 - \beta_0^3 \eta^3 \bigg) a_s^4 \nonumber \\ 
      &&   +\; \left(-\beta_3\eta  + \frac{3}{2} \beta_1^2\eta^2 + 3 \beta_0 \beta_2\eta^2  
		      - \frac{13}{3} \beta_0^2 \beta_1 \eta^3
                	+ \beta_0^4 \eta^4
				\right) a_s^5 \\ 
      && +\; \bigg( -\beta_4\eta + \frac{7}{2} \beta_1 \beta_2\eta^2 
		          + \frac{7}{2} \beta_0 \beta_3\eta^2  
						- \frac{35}{6} \beta_0 \beta_1^2 \eta^3 
               				- 6 \beta_0^2 \beta_2 \eta^3 
                           + \frac{77}{12} \beta_0^3 \beta_1 \eta^4
       			- \beta_0^5 \eta^5
				\bigg)a_s^6 
		+\mathcal{O}(\eta^6;a_s^7)~.  \nonumber 
\eeqn
Here the series has been reordered in powers of $a_s\equiv a_s(s_0)$ and 
we use the RGE $\beta$-function\footnote
{The full expressions for an arbitrary number of quark flavours ($n_f$) 
   are: $\beta_0 = \frac{1}{4}\left(11 - \frac{2}{3}n_f\right)$,
	 $\beta_1 = \frac{1}{16}\left(102 - \frac{38}{3}n_f\right)$,
	 $\beta_2 = \frac{1}{64}\left(\frac{2857}{2} - \frac{5033}{18}n_f 
     + \frac{325}{54}n_f^2\right)$, and
	 $\beta_3 = \frac{1}{256}\left[
	       \frac{149753}{6} + 3564\,\zeta_3 -
	       \left(\frac{1078361}{162} + \frac{6508}{27}\,\zeta_3\right)n_f \right. \,+\,
          \left. \left(\frac{50065}{162} + \frac{6472}{81}\,\zeta_3\right)n_f^2 +
	       \frac{1093}{729}n_f^3\right]$,
	where the $\zeta_{i=\{3,4,5\}}=\{1.2020569,\pi^4/90,1.0369278\}$ are the Riemann 
	$\zeta$-functions. The $\beta_{n\ge 4}$ are unknown.
} 
as defined in~\cite{rit}.
\pskip
Computing the contour 
integral~(\ref{eq:an}), and ordering the contributions according to their 
powers in $a_s$, leads to the familiar expression for fixed-order perturbation 
theory (FOPT)~\cite{pert}
\beq 
\label{eq:kngn}
   \delta^{(0)} = 
       \sum_{n=1}^\infty \left[\tilde{K}_n(\xi) + g_n(\xi)\right]
       a_s^n(\xi s_0)\,,
\eeq
where the $g_n$ are functions of $\tilde{K}_{m<n}$ and $\beta_{m<n-1}$, and of
elementary integrals with logarithms of power $m<n$ in the integrand. Setting 
$\xi=1$ and replacing all known $\beta_i$ coefficients by their numerical values 
for $n_f=3$ gives~\cite{rmp,kataev}
\beqn
\label{eq:delta0exp}
   \delta^{(0)}
   &=&
      a_s(s_0)
      + (K_2 +  3.5625)\,a_s^{2}(s_0)
      + (K_3  + 19.995)\,a_s^{3}(s_0) 
      + (K_4 + 78.003)\,a_s^{4}(s_0) \\
   &&
      +\: (K_5 + 307.787)\,a_s^{5}(s_0)
      + (K_6 + 17.813\,K_5  
      + 1.5833\,\beta_4 - 5848.19)\,a_s^{6}(s_0)\:,\nonumber
\eeqn
where for the purpose of later studies we have kept terms up to \sixth order.
\pskip
The FOPT series is truncated at a given order despite the fact that parts of 
the higher coefficients $g_{n>4}(\xi)$ are known and could be 
resummed: these are the higher order terms of the $a_s(s)$ expansion that 
are functions of $\beta_{n\le3}$ and $K_{n\le4}$ only. Moreover, at each 
integration step, the expansion~(\ref{eq:astaylor}) with respect to 
the physical value $a_s(s_0)$ is used to predict $a_s(s)$ on the entire $|s|= s_0$ 
contour. This might not always be justified, and leads to systematic errors
as discussed in Sec.~\ref{sec:ptcomp}.
\pskip
A more accurate approach to the solution of the contour integral~(\ref{eq:an}) 
is to perform a direct numerical evaluation by step-wise integration.
At each integration step, it takes as input for the running $a_s(s)$ the solution 
of the RGE to four loops, computed using the value from the previous 
step~\cite{pivov,pert}. It implicitly provides a partial resummation of the 
(known) higher order logarithmic contributions, and does not require the validity 
of the $a_s(s)$ Taylor series for large absolute values of the expansion 
parameter $\eta$.
This numerical solution of Eq.~(\ref{eq:knan}) is referred to as 
{\it contour-improved} perturbation theory (CIPT).

\subsubsection{Alternative Perturbative Expansions}
\label{sec:ecpt}

Inspired by the pioneering work in~\cite{grunberg1,grunberg2,dhar1,dhar2,pms} 
the {\em effective charge approach} to the perturbative prediction
of \Rtau (ECPT) has triggered many studies~\cite{maxwellrs1,raczka,pivoveffch}. 
The advocated advantage of this technique is that the perturbative 
prediction of the effective charge is renormalisation scheme  and
scale invariant since it is a physical observable. The effective $\tau$ 
charge is defined by $a_\tau = \delta^{(0)}$. 
The ECPT scheme has been used in the past to estimate the unknown
higher-order perturbative coefficient $K_4$, by exploiting the mediocre convergence
of the series (because $a_\tau(\mtau^2)\simeq1.8\cdot a_s(\mtau^2)$). As 
pointed out in~\cite{chetkuehn}, these estimates missed the actual value of $K_4$ 
by approximately a factor of two. One reason for this disagreement 
may come from the fact that these methods 
neglected the contributions from the next higher and also unknown orders. Owing to the 
insufficient convergence, the uncertainty on the coefficient estimate introduced by this
neglect is significant and exceeds the errors quoted~\cite{rmp}.
\pskip
For completeness we also mention the {\em large-$\beta_0$ expansion}, which is an 
approximation to the full FOPT result assuming the dominance of the $[\beta_0 a_s(-s)]^n$ 
term. It is thus possible to derive estimates for the FOPT coefficients of a given 
perturbative series at all orders by neglecting higher order terms in the $\beta$-function. 
The  large-$\beta_0$ expansion corresponds to inserting
chains of fermion loops into the gluon propagators and to determining 
the impact on the quark-antiquark vacuum polarisation. The procedure provides hence a 
{\it naive non-abelianisation} of the theory, because the lowest-order radiative corrections 
do not include gluon self-coupling. 
As an illustration, the $\Rtau$ FOPT series~(\ref{eq:kngn}) can be expanded as
$
    \delta^{(0)}(s) = a_s\sum_{n=0}a_s^n\left(d_n\beta_0^{n} 
                                              + \delta_n\right)\:,
$
where $d_n\beta_0^{n}+\delta_n=K_{n+1}+g_{n+1}$ (setting $\xi=1$).
The coefficients $d_n$ are computed in terms of fermion bubble 
diagrams~\cite{broadhurst}, where they are identified with their leading-$n_f$ 
pieces $d_n^{[n]}$ in the expression $d_n = d_n^{[n]}n_f^k + \dots + d_k^{[0]}$.
Neglecting the corrections $\delta_n$, the above series
leads to the large-$\beta_0$ expansion of $\delta^{(0)}$. 
The first elements of the series are~\cite{benekebraun}:
$      d_0         = 1$,
$      d_1\beta_0   = 5.119$,
$      d_2\beta_0^2 = 28.78$,
$      d_3\beta_0^3 = 156.6$,
$      d_4\beta_0^4 = 900.9$,
$      d_5\beta_0^5 = 4867$.
They compare reasonably well with the FOPT terms~(\ref{eq:delta0exp})
where these are known, in particular the large size of the \fourth-order
term has been anticipated ($K_4\sim79$). However, it turns out that the estimated 
coefficients of the Adler series itself (before integration on the contour) do not 
compare well with the exact solutions, which emphasises the uncontrolled theoretical 
uncertainties associated with this method~\cite{rmp}. 

\subsubsection{Comparing Perturbative Methods}
\label{sec:ptcomp}

This section updates and completes the discussion given in Secs.~3 and 8 of~\cite{rmp}, 
including here the known value of the \fourth-order perturbative coefficient
in the Adler function, $K_4$~\cite{chetkuehn}.  We perform a numerical
study of the FOPT and CIPT approaches to expose the differences between these two 
methods. Both use the Taylor series~(\ref{eq:astaylor}), and they assume that one 
can perform an analytic continuation of the solution of the RGE for complex values 
of $s$,\footnote
{One of the first limits of this hypothesis shows up in the discontinuity of the 
   imaginary part of $\as$ at $\phi = \pm \pi$, which is due to the cut of the 
   logarithm in the complex plane. 
}
namely along the circular contour of integration in Eq.~(\ref{eq:an}). One should 
thus make sure that the series is used only inside the domain of good convergence. As 
one approaches the limit of this domain, the error induced by the finite Taylor series 
increases. For CIPT the convergence is guaranteed because the integration
proceeds along infinitesimal steps such that $|\eta|\ll1$ everywhere. 
The situation is more complicated for FOPT as the absolute value of $\eta$
in Eq.~(\ref{eq:astaylor}) approaches $\pi$ close to the branch cut. 
\pskip
The tests carried out here use the expansion~(\ref{eq:astaylor}) to \sixth order 
in $a_s(s_0)$ (hence \fifth order in $\eta=\ln(s/s_0)$) --- if not stated otherwise, 
with estimates for $K_{5,6}$ and $\beta_4$ assuming a geometric growth of the 
corresponding series (\ie, $K_{5(6)}=K_{4(5)}(K_{4(5)}/K_{3(4)})$ and 
$\beta_4=\beta_3(\beta_3/\beta_2)$), and setting all coefficients at 
higher-orders than these to zero.

\subsubsection*{Taylor Series}

\begin{figure}[t]
\centerline{\includegraphics[width=0.49\columnwidth]{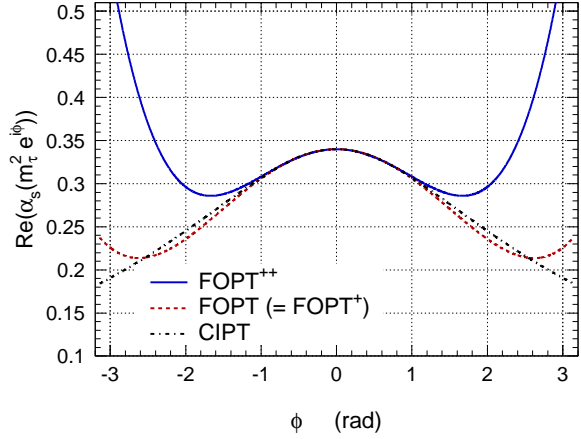}}
\vspace{-0.1cm}
\caption{Real part of \ass computed along the 
         $|s|=s_0$ contour for $\xi=1$, using respectively \FOPTpp (solid line, 
         see text), FOPT and \FOPTp (dashed, see text) and CIPT (dashed-dotted). } 
\label{fig:reAlphasInt}
\end{figure}

\begin{figure}[t]
\centerline{
\includegraphics[width=0.49\columnwidth]{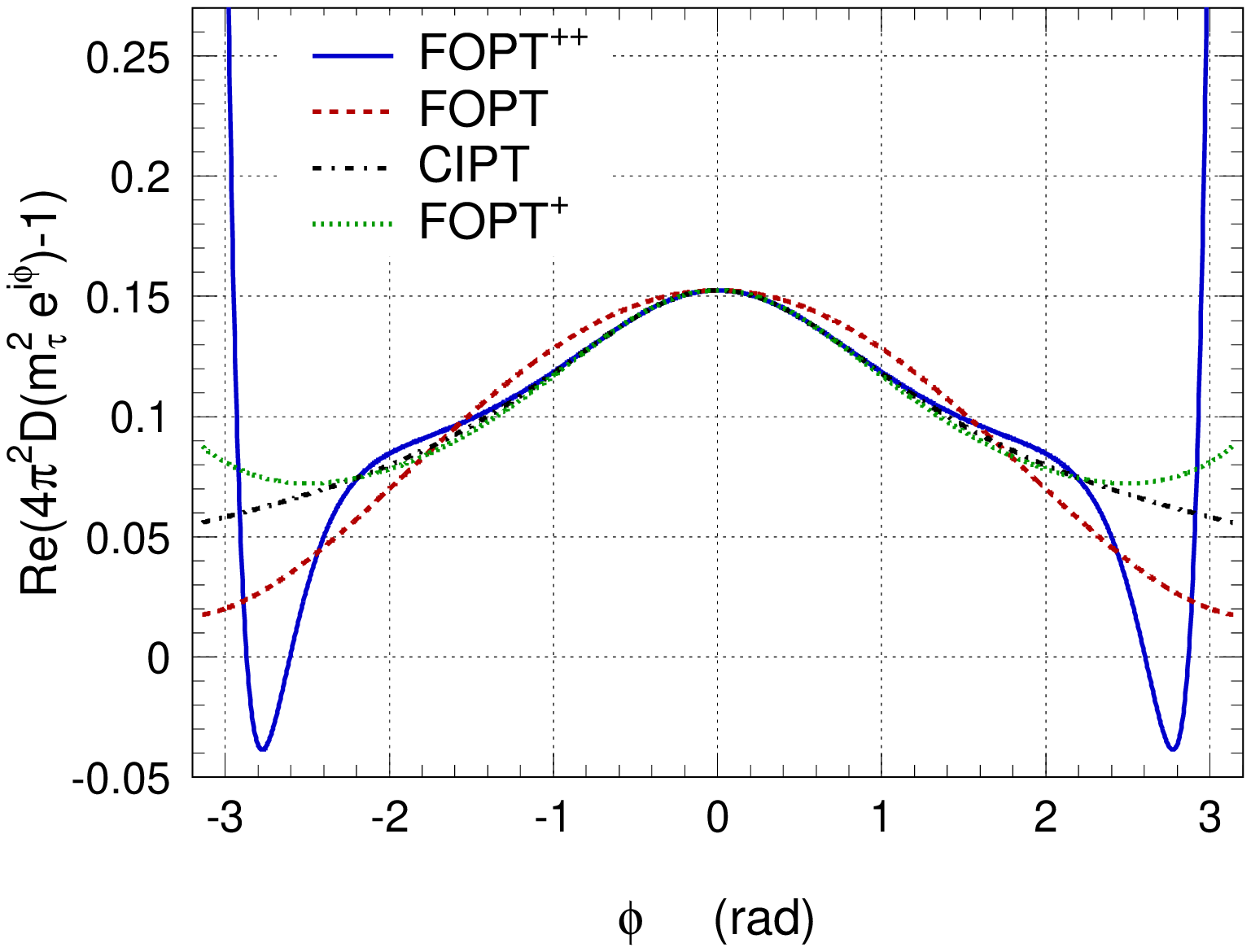}\hspace{0.3cm}
\includegraphics[width=0.49\columnwidth]{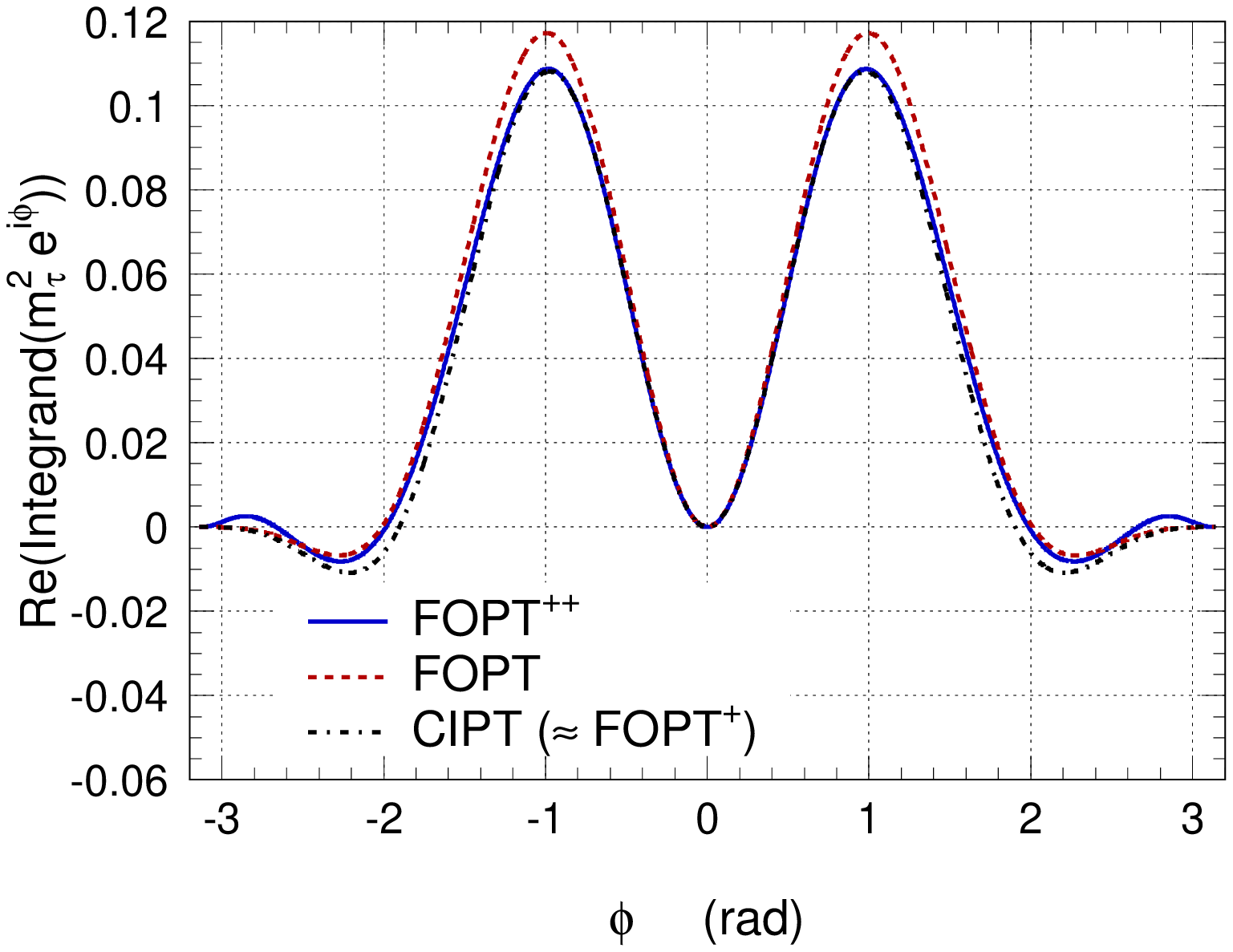}}
\vspace{-0.1cm}
\caption{Real part of $(4\pi^2 D(s) - 1)$ (left) and of the integrand in Eqs.~(\ref{eq:an}) and 
         (\ref{eq:knan}) (right), computed along the integration contour for $\xi=1$, 
         using respectively \FOPTpp (solid line), FOPT (dashed), CIPT 
         (dashed-dotted) and \FOPTp (dotted, not shown on the right hand plot 
         because it is almost indistinguishable from CIPT).} 
\label{fig:reAdler}
\end{figure}

To check the stability of the results obtained with FOPT, we consider a variant 
(denoted \FOPTpp) where all known or estimated terms of order $\eta^{n\le5}$ are kept
(\ie, including the known expressions with powers $a_s^{n=7}(s_0)$ and beyond),
which should reduce the error associated with the use of the Taylor expansion in FOPT. 
Figure~\ref{fig:reAlphasInt} shows the evolution of the 
real part of \ass along the integration circle as found for CIPT, FOPT and \FOPTpp. 
As expected, the values for CIPT and \FOPTmaybepp agree in the region around 
$\phi=0$ (the fix-point of the expansion in \FOPTmaybepp), but 
significant discrepancies occur elsewhere. For \FOPTpp we find large values 
for $\Re(\as)$ close to the branch cut. Estimating the convergence speed of 
the $\eta$ series~(\ref{eq:astaylor}) reveals that it is slower
for \FOPTpp, where larger powers of $a_s$ are kept, than for FOPT, for which
the series is truncated at $a_s^6$. Including higher orders $\eta^{n>5}$ in 
\FOPTpp we find that these terms dominate the value of $\Re(\as)$ near the branch 
cut, leading to large deviations from the correct evolution, which rise with 
the order $n$. On the contrary, for CIPT performed with infinitesimal integration 
steps, the full five-loop RGE solution is equivalent to Eq.~(\ref{eq:astaylor}),
\ie, ${\rm CIPT}=\CIPTpp$.\footnote
{To understand this feature, one can compare the errors induced by the Taylor 
   approximation for the FOPT and CIPT numerical procedures along the
   circular contour. To compute the contour integral, $N\gg1$ equidistant integration
   points along the contour are added.
   At the $j^{\rm th}$ point, the error on the value of $\as$ is given directly by 
   Eq.~(\ref{eq:astaylor}) for FOPT, whereas one can easily show that it is reduced 
   by the factor $j/N^{n+1}$ for CIPT, where $n=5$ is the expansion order in $\eta$. 
   Therefore, the error on the contour integral coming from the determination of $\as$ 
   is suppressed by $1/N^n$ in the case of CIPT compared to FOPT. 
}
\pskip
Although the values of $\as$ differ significantly on half of the integration domain, 
the standard FOPT and CIPT methods give similar results for the integral. This is 
because the integration kernel~(\ref{eq:rtauadler}) vanishes for $s=-s_0$ ($\phi=\pm \pi$),
suppressing the contributions to the integral coming from the region near the branch 
cut.\footnote
{In addition, a significant cancellation takes place in
   this region: for FOPT, the contribution of the contour integral vanishes
   on the intervals $[-\pi;-1.73]$ and $[1.73;\pi]$,
   whereas for CIPT a vanishing contribution comes from $[-\pi;-1.57]$ and $[1.57;\pi]$.
}
The main difference between the two results stems from the regions $\phi \approx \pm 2.1$ 
and $\phi \approx \pm 1$ (\cf left-hand plot of Fig.~\ref{fig:reAdler}).
In the region $|\phi| < 1$, the values of \ass estimated by 
the two methods are close, and the difference between the two integrands can be
ascribed to the truncation at the \sixth order in $a_s(s_0)$ for the integrand of 
FOPT.

\subsubsection*{Fixed-order Truncation}

In addition to employing a Taylor series in a region with questionable convergence 
properties, FOPT truncates the full expression of the contour integral in Eq.~(\ref{eq:kngn}). 
To disentangle the impact of these two approximations, we have tested another variant 
of FOPT (denoted \FOPTp), where Eq.~(\ref{eq:astaylor}) is used as is, but 
without truncating the Adler function (or equivalently $\delta^{(0)}$) at the 
\sixth order in $a_s(s_0)$. This method leads to a similar integrand as in CIPT,
with however the usual difference in the evolution. The left-hand plot of 
Fig.~\ref{fig:reAdler} 
shows the evolution of the real part of $(4\pi^2 D(s) - 1)$ along the contour for all methods. 
\FOPTp and CIPT differ close to the branch cut as a consequence of the deficient
Taylor approximation, with however little difference in the integration result~\cite{rmp}
due to the suppression by the integration kernel. The \FOPTpp approach without 
truncating the Adler function leads to a $\delta^{(0)}$ that lies between CIPT 
and FOPT, with however unstable numerical dependence on the largest power in 
$\eta$ kept in the Taylor series. 

\subsubsection*{Numerical Comparisons}

\begin{figure}[t]
\centerline{
\includegraphics[width=0.49\columnwidth]{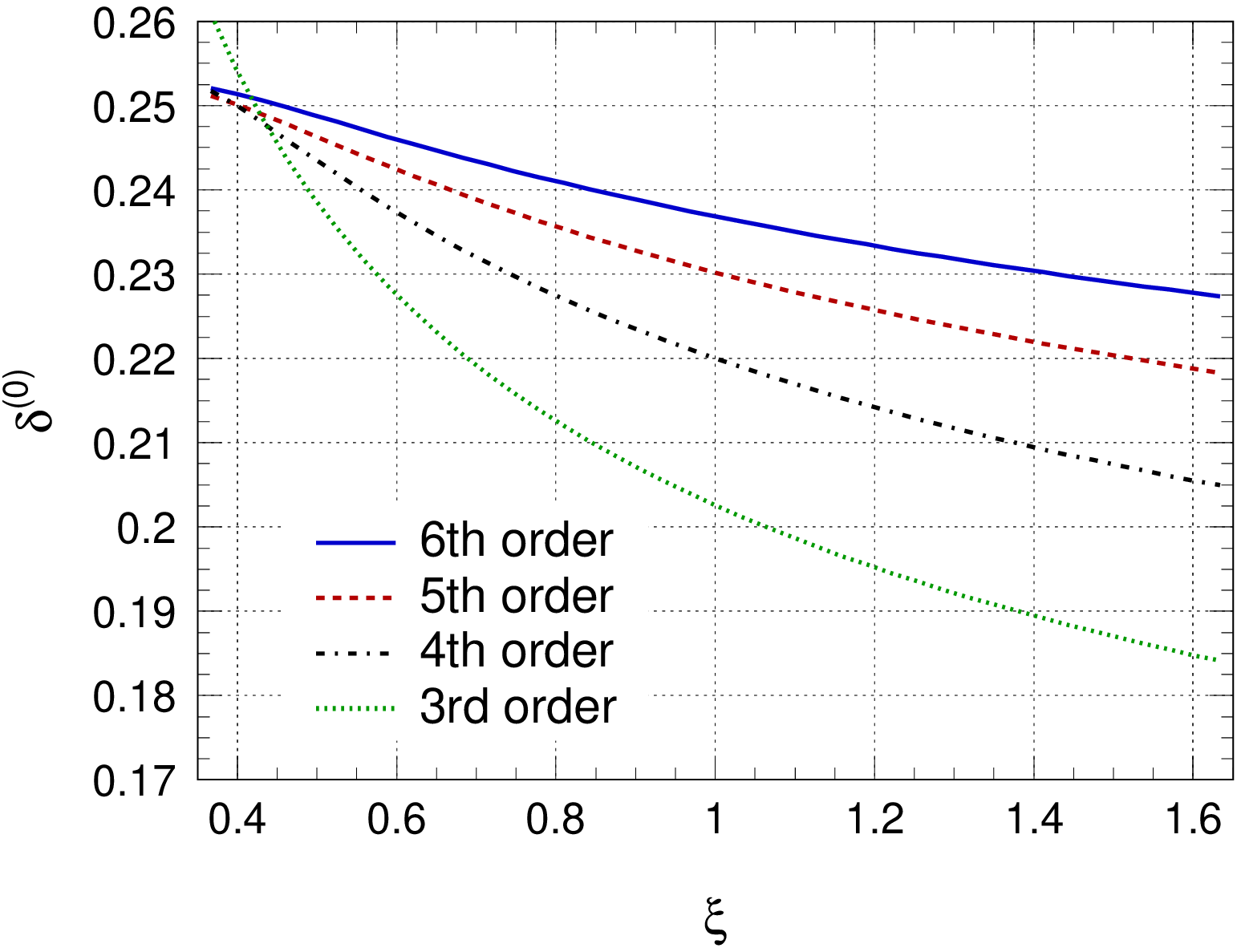}\hspace{0.3cm}
\includegraphics[width=0.49\columnwidth]{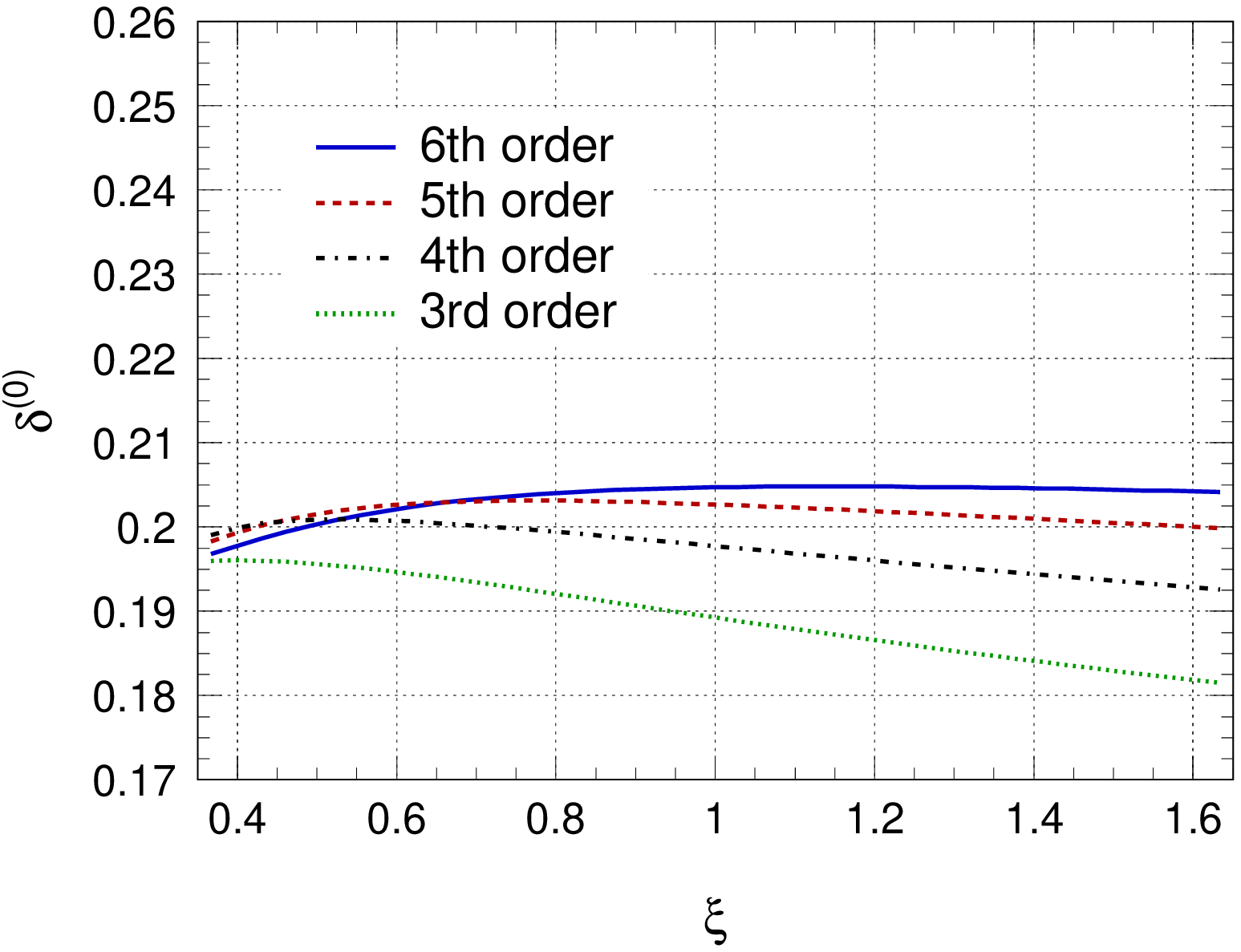}}
\vspace{-0.1cm}
\caption{Scale dependence of $\delta^{(0)}$ in \Rtau computed at the \third to the 
         estimated \sixth order with FOPT (left) and CIPT (right).} 
\label{fig:d0Tau}
\end{figure}

\begin{table*}[t]
  \caption[.]{\label{tab:intsol}
              Massless perturbative contribution $\delta^{(0)}$ in \Rtau using FOPT, CIPT 
              and the large-$\beta_0$ expansion, respectively, and computed for 
              $\asTau=0.34$. The unknown higher-order $K_{5,6}$ and $\beta_{4}$ 
              coefficients are estimated by assuming a geometric growth (see text), 
              while the remaining ones are set to zero. The quoted uncertainties $\delta$ 
              correspond to the indicated error ranges.}
\begin{center}
\setlength{\tabcolsep}{0.0pc}
\begin{tabular*}{\textwidth}{@{\extracolsep{\fill}}lrrrrrrrrr} 
\hline\noalign{\smallskip}
Pert. Method	
        & $n=1$	 & $n=2$  & $n=3$   &  $n=4$  & $(n=5)$ & $(n=6)$   & $\sum_{n=1}^4$ & $\sum_{n=1}^5$ & $\sum_{n=1}^6$ \\
\noalign{\smallskip}\hline\noalign{\smallskip}
FOPT ($\xi=1$)
	& 0.1082 & 0.0609 & 0.0334  & 0.0174  & 0.0101  & 0.0067 & 0.2200 & 0.2302 & 0.2369 \\
$\delta(\beta_4 \pm 100\%)$ 
	& 0 & 0 & 0 & 0 & 0 & $\pm 0.0006$ & 0 & 0 & $\pm 0.0006$ \\
$\delta(K_5 \pm 100\%)$ 
	& 0 & 0 & 0 & 0 & $\pm 0.0056$ & $\pm 0.0108$ & 0 & $\pm 0.0056$ & $\pm 0.0164$ \\
$\delta(K_6 \pm 100\%)$ 
	& 0 & 0 & 0 & 0 & 0 & $\pm 0.0047$ & 0 & 0 &$\pm 0.0047$ \\
$\delta(\xi \pm 0.63)$ 
	& -- & -- & -- & -- & -- & -- & $^{+0.0317}_{-0.0151}$ & $^{+0.0209}_{-0.0119}$ & $ ^{+0.0152}_{-0.0095}$ \\
\noalign{\smallskip}\hline\noalign{\smallskip}
CIPT ($\xi=1$)           	
	& 0.1476 & 0.0295 & 0.0121 & 0.0085  & 0.0049  &  0.0020   &  0.1977  & 0.2027 & 0.2047 \\
$\delta(\beta_4 \pm 100\%)$ 
	& $\mp 0.0003$ & $\mp 0.0001$ & $\mp 0.0001$ & $\mp 0.0001$ & $\mp 0.0001$ & $\mp 0.0001 $ & $\mp 0.0006$ & $\mp 0.0007$ & $\mp 0.0008$ \\
$\delta(K_5 \pm 100\%)$ 
	& 0 & 0 & 0 & 0 & $\pm 0.0049 $  & 0  & 0 & $\pm 0.0049$ & $\pm 0.0049 $ \\
$\delta(K_6 \pm 100\%)$ 
	& 0 & 0 & 0 & 0 & 0 & $\pm 0.0020$ & 0 & 0 & $\pm 0.0020$ \\
$\delta(\xi \pm 0.63)$ 
	& -- & -- & -- & -- & -- & -- & $^{+0.0032}_{-0.0051}$ & $^{+0.0005}_{-0.0044}$ & $ ^{+0.0001}_{-0.0079}$ \\
\noalign{\smallskip}\hline\noalign{\smallskip}
Large-$\beta_0$ expansion
	& 0.1082 & 0.0600 & 0.0364  & 0.0215  & 0.0134  & 0.0078  & 0.2261 & 0.2395  & 0.2473 \\
\noalign{\smallskip}\hline
\end{tabular*}
  \end{center}
\end{table*}
Table~\ref{tab:intsol} summarises the contributions of the orders $n\le6$ in PT 
to $\delta^{(0)}$ for FOPT, CIPT and the large-$\beta_0$ expansion,\footnote
{We do not include ECPT into the present study, because --- as concluded
  in~\cite{rmp} --- the convergence of the perturbative series is insufficient 
  for a precision determination of \asTau.
} 
using as benchmark value $\asTau=0.34$, and $\xi=1$. For systematic studies 
we vary $\xi$ in the range $\xi\cdot m_{\tau}^2=m_{\tau}^2 \pm 2\gev^2$, and
the maximum observed deviations with respect to $\xi=1$ are reported in the 
corresponding lines of Table~\ref{tab:intsol}. 
We assume a geometric growth of the perturbative terms for all unknown PT and RGE 
coefficients, with 100\% uncertainty assigned to each of them for the purpose of 
illustration. We recall that the \nth contributions to the FOPT 
and CIPT series should be compared with care. Whereas the FOPT contributions can be 
directly obtained from Eq.~(\ref{eq:kngn}), the entanglement of the different perturbative 
orders generated by CIPT prevents us from separating the contributions in powers of 
$a_s(s_0)$. Instead, the columns given for CIPT in Table~\ref{tab:intsol} 
correspond to the terms in Eq.~(\ref{eq:knan}). If the two methods were 
equally well suited for the integration, their column sums should converge to 
the same value.
\pskip
The variations of $\delta^{(0)}$ with the scale parameter $\xi$ 
are strongly non-linear (\cf the asymmetric errors in Table~\ref{tab:intsol} 
and the functional forms plotted for FOPT (left) and CIPT (right) 
in Fig.~\ref{fig:d0Tau}). CIPT exhibits significantly less renormalisation 
scale dependence than FOPT at order $n=4$, while the interpretation of 
the subsequent orders strongly depends on the values used for the unknown 
coefficients $K_{n\ge5}$.

\subsubsection*{Conclusions}

\begin{figure}[t]
\vspace{0.5cm}
\centerline{
\includegraphics[width=0.49\columnwidth]{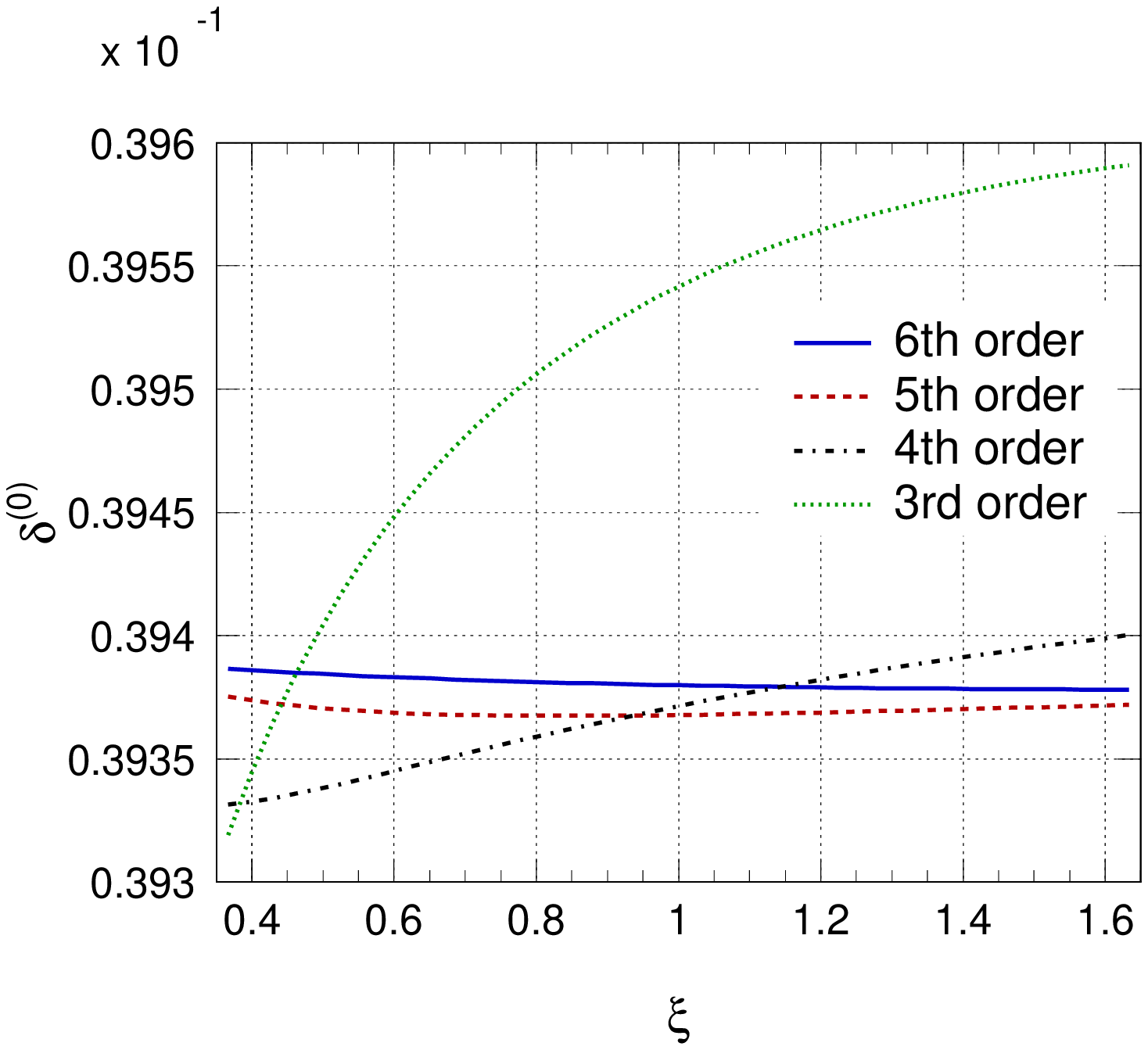} \hspace{0.3cm}
\includegraphics[width=0.49\columnwidth]{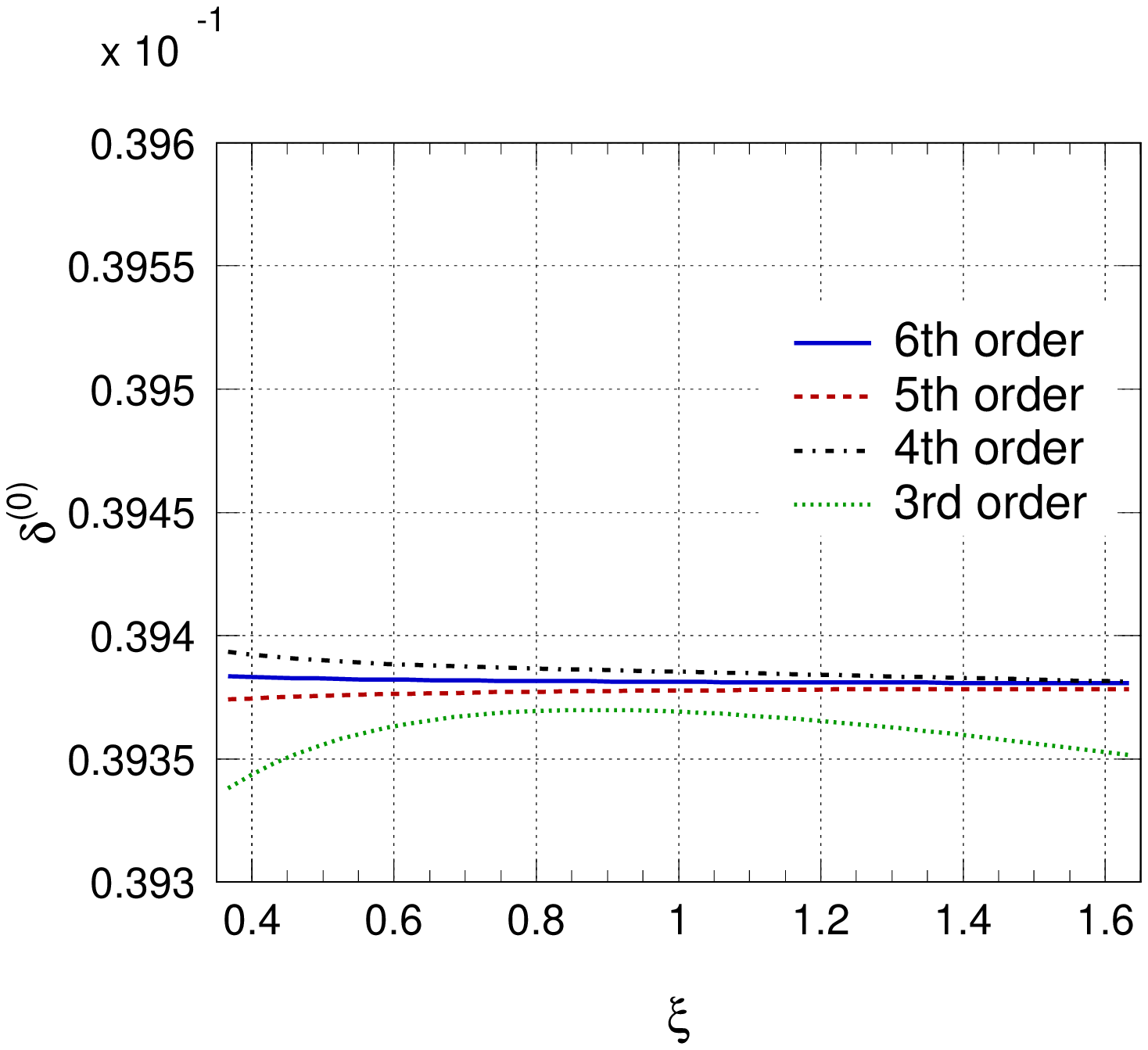}}
\vspace{-0.1cm}
\caption{Scale dependence of $\delta^{(0)}$ in $\Ree^{(5)}(\mZ^2)$ computed at the 
         \third to the estimated \sixth order with FOPT (left) and CIPT (right).} 
\label{fig:d0e+e-}
\end{figure}
The CIPT series is found to be better behaved than FOPT and is therefore to be 
preferred for the numerical analysis of the $\tau$ hadronic width. This preference 
is also supported by the analysis of the integrand in the previous section, 
suggesting a pathological behaviour of FOPT for $a_s$ 
near the branch cut. Our coarse extrapolation of the higher-order coefficients 
could indicate that minimal sensitivity is reached at $n\sim5$ for FOPT, while the 
series further converges for CIPT. The uncertainties due to $K_5$ and $K_6$ are 
smaller for CIPT whereas the one due to the unknown value of $\beta_4$ is similar 
in both approaches.
The difference in the result observed when using a Taylor 
expansion and truncating the perturbative series after integrating
along the contour (FOPT) with the exact result at given order (CIPT) 
exemplifies the incompleteness of the perturbative series. The situation 
is even worse since, not only large known contributions are neglected
in FOPT, but the series is also used in a domain where its convergence is not 
guaranteed: taking the difference between CIPT and FOPT as an estimate of
the related systematic error overestimates the uncertainty due to the 
truncation of the perturbative series. In the line of this discussion, and 
following~\cite{rmp}, we will not use this prescription to estimate the systematic 
error on the truncation of the series, and we will limit the analysis to the 
uncertainties coming from the study of CIPT only.
\pskip
The discrepancies found between FOPT and CIPT at $|s|=m_\tau^2$ 
are reduced drastically when computing $\Ree^{(5)}(\mZ^2)$ (see Fig.~\ref{fig:d0e+e-} 
and Table~\ref{tab:intsol_z}). The small value of \asZ ensures 
a much better convergence of the perturbative series. The better convergence also 
leads to a tiny scale dependence, which is even smaller for CIPT than for FOPT, and
hence to small theoretical uncertainties.
\pskip
\begin{table*}[t]
  \caption[.]{\label{tab:intsol_z}
              Massless perturbative contributions to $\delta^{(0)}$
              in $\Ree^{(5)}(\mZ^2)$ using FOPT and CIPT, respectively, and computed 
              for $\asZ=0.12$. The unknown higher-order $K_{5,6}$ and $\beta_{4}$ 
              coefficients are estimated by assuming a geometric growth, 
              while the others are set to 0. The quoted uncertainties $\delta$ 
              stem from the indicated range of values for the unknown parameters 
              and from the renormalisation scale.}
\begin{center}
\setlength{\tabcolsep}{0.0pc}
\begin{tabular*}{\textwidth}{@{\extracolsep{\fill}}lrrrrrrrrr} 
\hline\noalign{\smallskip}
Pert. Method	
        & $n=1$	 & $n=2$  & $n=3$   &  $n=4$  & $(n=5)$ & $(n=6)$   & $\sum_{n=1}^4$ & $\sum_{n=1}^5$ & $\sum_{n=1}^6$ \\
\noalign{\smallskip}\hline\noalign{\smallskip}
FOPT ($\xi=1$)
	& $0.038197$ & $0.002056$ & $-0.000712$  & $-0.000170$  & $-0.000004$  & $0.000012$ & $0.039372$ & $0.039368$ & $0.039380$ \\
$\delta(\beta_4 \pm 100\%)$ 
	& 0 & 0 & 0 & 0 & 0 & 0 & 0 & 0 &  0 \\
$\delta(K_5 \pm 100\%)$ 
	& 0 & 0 & 0 & 0 & $\mp 10^{-5} $ & 0 & 0 & $\mp 10^{-5} $ & $\mp 10^{-5} $ \\
$\delta(K_6 \pm 100\%)$ 
	& 0 & 0 & 0 & 0 & 0 & $\pm 5 \cdot 10^{-6}$ & 0 & 0 & $\pm 5 \cdot 10^{-6}$ \\
$\delta(\xi \pm 0.63)$ 
	& -- & -- & -- & -- & -- & -- & $ ^{+29}_{-40}\cdot 10^{-6} $ & $ ^{+7.4}_{-0.3}\cdot 10^{-6} $ & $ ^{+6.7}_{-1.9}\cdot 10^{-6} $ \\
\noalign{\smallskip}\hline\noalign{\smallskip}
CIPT ($\xi=1$)           	
	& $0.037462$ & $0.001941$ & $-0.000034$ & $0.000016$ & $-0.000008$ & $0.000003$  & $0.039385$ & $0.039378$  & $0.039381$ \\
$\delta(\beta_4 \pm 100\%)$ 
	& $<10^{-6}$ & $\approx 0$ & $\approx 0$ & $\approx 0$ & $\approx 0$ &  $\approx 0 $ & $<10^{-6}$ & $<10^{-6}$ & $<10^{-6}$ \\
$\delta(K_5 \pm 100\%)$ 
	& 0 & 0 & 0 & 0 & $\mp 8 \cdot 10^{-6} $  & 0  & 0 & $\mp 8 \cdot 10^{-6}$ & $\mp 8 \cdot 10^{-6}$ \\
$\delta(K_6 \pm 100\%)$ 
	& 0 & 0 & 0 & 0 & 0 & $\pm 3 \cdot 10^{-6} $ & 0 & 0 & $\pm 3 \cdot 10^{-6}$ \\
$\delta(\xi \pm 0.63)$ 
	& -- & -- & -- & -- & -- & -- & $^{+8.2}_{-4.1}\cdot10^{-6}$ & $ ^{+0.6}_{-3.7}\cdot 10^{-6}$ & $ ^{+2.3}_{-0.5}\cdot 10^{-6}$ \\
\noalign{\smallskip}\hline
\end{tabular*}
  \end{center}
\end{table*}

\subsection{Quark-Mass and Nonperturbative Contributions}
\label{sec:nonpert}

Following SVZ~\cite{svz}, the first contribution to \Rtau beyond the 
$D=0$ perturbative expansion is the non-dynamical quark-mass correction 
of dimension $D=2$, \ie, corrections  scaling like $1/\mtau^2$. 
The leading $D=2$ corrections induced by the light-quark masses are
computed using the running quark masses evaluated at the two-loop level
(denoted $\overline m$ in the following).
The evaluation of the contour integral in FOPT~\cite{bnp} leads to terms
$\delta^{(2,m_q)}_{ud,V/A} \propto \overline m_{u,d}^2(\mtau^2)/\mtau^2,\;
\overline m_{u}(\mtau^2)\overline m_{d}(\mtau^2)/\mtau^2$, which are small.
\pskip
The dimension $D=4$ operators have dynamical contributions from the gluon 
condensate $\GG$ and the light $u,d$ quark condensates 
$\langle m_iq_iq_i\rangle$, which are the vacuum expectation values of the gluon field 
strength-squared and of the scalar quark densities, respectively. The remaining $D=4$ 
operators involve the running quark masses to the \fourth power. Solving the contour 
integral~\cite{bnp} results in terms $\delta_{ud,V/A}^{(4)}\propto
\as^2(\mtau^2)\GG/\mtau^4,\;\langle m_q \qbar q\rangle/\mtau^4,\;
{\cal O}_4(\overline m_q^4/\mtau^4)$, where remarkably the contribution from the 
gluon condensate vanishes at the first order in \asTau. 
\pskip
The contributions from dimension $D=6$ operators are more delicate to analyse. 
The most important operators arise from four-quark terms of 
the form $\qbar_i\Gamma_1q_j\qbar_k\Gamma_2q_l$. We neglect other operators, such 
as the triple gluon condensate whose Wilson coefficient vanishes at order 
\as, or those which are suppressed by powers of quark masses, 
in the evaluation of the contour integrals performed in~\cite{bnp}.
The large number of independent operators of the four-quark type occurring 
in the $D=6$ term can be reduced by means of the
{\em vacuum saturation} assumption~\cite{svz}. 
The operators are then expressed as products of (two-)quark 
condensates $\asmu\langle\qbar_iq_i(\mu)\rangle\langle\qbar_jq_j(\mu)\rangle$.
Since the scale dependence of the four-quark and two-quark operators are different,
such factorisation can hold for a specific value of the renormalisation scale (at best). 
To take into account this problem as well as likely deviations from the vacuum saturation 
assumption, one can introduce an effective parameter $\rho$ (in principle scale-dependent) 
to replace the four-quark contribution by $\rho \as\langle\qbar q\rangle^2$. The 
effective $D=6$ term obtained in this way is~\cite{bnp} 
$\delta_{ud,V/A}^{(6)}\propto \rho\as\langle\qbar q\rangle^2/\mtau^6$, with 
a relative factor of $-7/11$ between vector and axial vector contributions.
\pskip
The $D=8$ contribution has a structure of non-trivial quark-quark, 
quark-gluon and four-gluon condensates whose explicit form is given 
in~\cite{dimeight}. For the theoretical prediction of \Rtau\  it is customary
to absorb the whole long- and short-distance parts into the scale 
invariant phenomenological $D=8$ operator $\langle{\cal O}_8\rangle$,
which is fit simultaneously with \as\  and the other unknown 
nonperturbative operators. Higher-order contributions from $D\ge10$ 
operators to \Rtau are expected to be small since, like in the case of the 
gluon condensate, constant terms and terms in leading order in \as vanish after 
integrating over the contour. We will not consider these terms in the following.

\subsection{Impact of Quark-Hadron Duality Violation}
\label{sec:dualityViolation}

A matter of concern for the QCD analysis at the $\tau$ mass scale is the 
reliability of the theoretical description, \ie, the use of the OPE to 
organise the perturbative and nonperturbative expansions, and the control
of unknown higher-order terms in these series. A reasonable stability test 
consists in varying $\mtau$ continuously to lower values $\sqrt{s_0}\le \mtau$ for 
both theoretical prediction and measurement, which is possible since the 
shape of the full $\tau$ spectral function is available. This test was
successfully carried out~\cite{aleph_asf,opal_vasf,rmp} and confirmed the validity 
of the approach down to $s_0 \sim 1\,\gev^2$ with an accuracy of 1--2\%.
In this section, we consider a different test of the sensitivity of the 
analysis to possible OPE violations.
\pskip
The SVZ expansion provides a description of the correlator $\Pi$ (or
of the Adler function $D$) for values of the incoming momentum in the deep 
Euclidean region, based on the separation between large and soft momenta 
flowing through the diagrams associated to this correlator.   
If the OPE description were accurate, we could check the cogency of 
this description by performing an analytic continuation of the OPE 
to any value of the momentum in the physical region and comparing it
with the spectral functions in  Fig.~\ref{fig:tauSF}. As 
seen from these figures, perturbative QCD describes the asymptotic
behaviour of the functions, but fails to reproduce their details.
\pskip
The OPE suffers from a similar failure as can be expected from the 
intrinsic nature of the OPE 
procedure~\cite{svz,quinnetal,braaten88,shifman,cataetal}: it only yields a truncated 
expansion in the first powers of $1/Q$, \ie, the singularities near $x=0$ of 
$\Pi^{\mu\nu}$ (\cf Eq.~(\ref{eq:correl_def})). Therefore, it misses 
singularities for finite $x^2$ or $x^2\to\infty$ related to long-distance 
effects. Even a large momentum $q$ flowing through the vacuum polarisation 
diagrams may be split into a soft quark-antiquark pair and soft 
gluons: this physical possibility cannot be properly described by OPE, since
no separation can be performed between hard and soft physics in such a situation.
One expects for some of these effects to yield terms proportional to $\exp(-\lambda Q)/Q^k$ 
or $\exp(-\lambda^2 Q^2)/Q^\l$ (where $k,\l$ are positive and $\lambda$ is 
a typical hadronic distance), which are exponentially suppressed in the deep 
Euclidean region and thus absent in the truncated OPE series. But once these 
terms are continued analytically along the branch cut, they generate a (power 
suppressed or exponentially suppressed) oscillatory behaviour of the spectral 
function, which is similar to the one in Fig.~\ref{fig:tauSF}. Such
a behaviour is generally called ``violation of local quark-hadron duality''. 
\pskip
To determine $R_\tau$, we compute the convolution of the OPE expression
of the Adler function with a kernel along the circle of radius $s_0$. We
know that duality violation will have a small impact for the two regions close
to the real axis (these terms are exponentially suppressed in the Euclidean region, 
and the kernel vanishes for $s=s_0$). But to assess the systematic uncertainties related
to the use of OPE, it is instructive --- even if very approximate --- to simulate
the contributions of duality violating terms on the rest of the circle. For this 
purpose, we use two different models proposed in~\cite{shifman}, which provide a 
coarse and rather qualitative description of such effects (one of these models has 
been very recently reconsidered in~\cite{cataetal2} to investigate duality-violating 
effects on the determination of nonperturbative condensates from ALEPH data in the 
vector channel). In both cases, one does 
not aim at a complete description of the correlator $\Pi$, but focuses on
the deviation between the full description and its truncated OPE expansion
$\Delta\Pi=\Pi-\Pi^{\rm OPE}$. In the first model $(I)$ the quarks propagate in an 
instanton background field with a fixed size $\rho$, leading to the duality violation
\beq
   \Delta\Pi^{(I)}(Q) = \frac{C_{I}}{Q^2} K_1(Q\rho) K_{-1}(Q\rho)\,,
\eeq
where the $K_{(-)1}$ are modified Bessel functions of the second kind.
The second model $(II)$ mimics a comb of resonances with a width that grows with the
energy, so that they overlap progressively when the energy increases
\beq
   \Pi^{(II)}(Q) = -\frac{1}{4\pi^2}\frac{1}{1-B/3\pi}
                  \left(\psi(z)+\frac{1}{z}\right)\,.
\eeq
Here $\psi(z)$ is the di-gamma function, and $z=(Q^2/\sigma^2)^{1-B/3\pi}$,
where $\sigma$ parametrises the offset between the resonances, and $B$ their
(growing) widths. In this model, one can define $\Pi^{\rm OPE}$ as the expansion
in powers of $1/z$ (up to $z^4$ here, since we neglect operators of $D=10$
and beyond). Duality violations are encoded in
$\Delta\Pi^{(II)}=C_{II}(\Pi-\Pi^{\rm OPE})^{(II)}$. The factors $C_{I,II}$ 
are normalisation constants.
\pskip
One can check that the two models share the same features: they are exponentially 
suppressed in the Euclidean region, and exhibit a branch cut for time-like values 
of $s$, such that they contribute to the spectral functions with oscillations 
decreasing in amplitude when the energy increases. They differ by the dependence 
of their oscillation frequency on the energy: the instanton model oscillates like 
$\sin(\sqrt{s}\rho)$, while the resonance model varies like $\sin(s/\sigma)$.
\pskip
To investigate the numerical impact of quark-duality violation on our
results, we vary for each model the parameters and fix the normalisation
such that the imaginary part of sum of the perturbative QCD computation 
and of the duality-violating terms match smoothly the $V+A$ spectral function
near $s=m_\tau^2$. We then compute the contribution of the duality-violating
part to $\delta^{(0)}$ by performing the contour integral (\ref{eq:rtauadler}).
For the instanton model we asymptotically reproduce the data
for $\rho$ values between 2.4 and $4.4\,\gev^{-1}$, leading to
a contribution to $\delta^{(0)}$ below $4.5 \cdot 10^{-3}$. For the
resonance model we find values for $\sigma^2$ between $1.65$ and $2\,\gev^2$,
and $B$ between 0.3 and 0.6, leading to a contribution to $\delta^{(0)}$ 
below $7\cdot 10^{-4}$. 
These limits are however quite conservative because the models used exhibit significant 
oscillations in the $V+A$ spectral function. Although allowed 
by the ALEPH data because of the larger error bars close to the $m_\tau^2$ endpoint,
such oscillations are disfavored by the overall pattern of the spectral function, with 
oscillation amplitudes that are strongly suppressed above $1\,\gev$.
Even though these two models could be improved in many ways, it is
hard to see how their contributions to $\delta^{(0)}$ could be enhanced by an order
of magnitude such that they would invalidate the OPE approach. At least in the case 
of the $V+A$ spectral function, we therefore expect the violation of quark-hadron duality 
to have a negligible impact on our results. In the next section, we will see that 
the induced error on $\delta^{(0)}$ remains well within the systematic uncertainties 
coming from other sources.

\section{Combined Fit}
\label{sec:spectralmoments}

Apart from the perturbative term, the full OPE contains contributions of nonperturbative
nature parametrised by higher-dimensional operators, whose value cannot be computed 
from first principles. It was shown in~\cite{pichledib} that one can exploit the shape 
of the spectral functions via weighted integrals to obtain additional constraints on 
\asTau and --- more importantly --- on the nonperturbative power terms. 

\subsection{Spectral Moments}

The {\em $\tau$ spectral moments} at $s_0=\mtau^2$ are defined by
\beq
\label{eq:moments}
   R_{\tau,V/A}^{k\l} =
       \intl_0^{\mtau^2} ds\,\left(1-\frac{s}{\mtau^2}\right)^{\!\!k}\!
                              \left(\frac{s}{\mtau^2}\right)^{\!\!\l}
       \frac{dR_{\tau,V/A}}{ds}\:,
\eeq
where $R_{\tau,V/A}^{00}=R_{\tau,V/A}$. Using the same argument of analyticity 
as for \Rtau, one can reexpress~(\ref{eq:moments}) as a contour integral along 
the circle $|s|=s_0$.  The factor $(1-s/\mtau^2)^k$ 
suppresses the integrand at $s=\mtau^2$ where the 
validity of the OPE is less certain and the experimental accuracy 
is statistically limited. Its counterpart $(s/\mtau^2)^\l$ projects upon
higher energies. The spectral information is used to fit simultaneously 
\asTau and the leading $D=4,6,8$ nonperturbative contributions.
Due to the intrinsic experimental correlations (all spectral moments rely 
on the same spectral function) only four moments are used as input to the fit.
\pskip
In analogy to \Rtau~(\ref{eq:delta}), the contributions to the moments originating 
from perturbative QCD and nonperturbative OPE terms are separated.
The prediction of the perturbative contribution takes the form
\beq 
   \delta^{(0,k\l)} = 
       \sum_{n=1}^\infty \tilde{K}_n(\xi) A^{(n,k\l)}(a_s)\:,
\eeq
with the functions~\cite{rmp}
\beqn
\label{eq:anmom}
A^{(n,k\l)}(a_s) 
   &=&
      \frac{1}{2\pi i}\hm\ointl_{|s|=\mtau^2}\hm\hm
      \frac{ds}{s}
       \Bigg[2\Gamma(3 + k)
             \bigg(\frac{\Gamma(1 + \l)}{\Gamma(4 + k + \l)}  
                   +\;2\frac{\Gamma(2 + \l)}{\Gamma(5 + k + \l)} 
             \bigg) \nonumber\\
   &&\hspace{2cm}
             -\; I\bigg(\frac{s}{s_0},1+\l,3+k\bigg) 
             - 2 I\bigg(\frac{s}{s_0},2+\l,3+k\bigg)
            \Bigg]
	     a_s^n(-\xi s)\:,
\eeqn
which make use of the elementary integrals 
$I(\gamma,a,b)=\int_0^\gamma t^{a-1}(1-t)^{b-1}dt$.
The contour integrals are numerically solved for the running 
$a_s(-\xi s)$ using the CIPT prescription.
\pskip
In the chiral limit and neglecting the small logarithmic $s$ dependence of the 
Wilson coefficients, the dimension $D$ nonperturbative contributions 
$\delta_{ud,V/A}^{(D,k\l)}$ to the spectral moments simplify greatly
(\cf matrix~(133) in~\cite{rmp}). One finds that with increasing weight $\l$ the 
contributions from low dimensional operators vanish. For example, the 
only nonperturbative contribution to the moment $R_{\tau,V/A}^{13}$ stems 
from dimension $D=8$ and beyond (neglected).
\pskip
For practical purpose it is more convenient to define moments that are normalised 
to the corresponding \RtauVA to decouple the normalisation from the shape of the 
$\tau$ spectral functions,
\beq
\label{eq:dkl}
   D_{\tau,V/A}^{k\l} =
     \frac{R_{\tau,V/A}^{k\l}}{R_{\tau,V/A}}~.
\eeq
The two sets of experimentally almost uncorrelated observables --- \RtauVA  
on one hand, and the moments $D_{\tau,V/A}^{k\l}$ on the other 
hand --- yield independent constraints on \asTau and thus provide an important 
test of consistency. The correlation between these observables is 
negligible in the $V+A$ case where \RtauVpA is calculated from the 
difference $R_\tau-R_{\tau,S}$, which is independent of the hadronic 
invariant mass spectrum. One experimentally obtains the $D_{\tau,V/A}^{k\l}$ 
by integrating weighted normalised invariant mass-squared spectra. 
The corresponding theoretical predictions are easily adapted. 
\pskip
The measured $V$, $A$ and $(V+A)$ spectral moments and their linear 
correlations matrices are given in Tables~\ref{tab:moments} and \ref{tab:momcorr}, 
respectively. Also shown are the central values of the theory prediction after
fit convergence (\cf Sec.~\ref{sec:fitResults}). The correlations 
between the moments are computed analytically from the contraction of the 
derivatives of two involved moments with the covariance matrices 
of the respective normalised invariant mass-squared spectra. In all cases, 
the negative sign for the correlations between the $k=1,\l=0$ and 
the $k=1,\l\ge1$ moments is due to the $\rho$ ($V$) and the $\pi$, $a_1$ ($A$) peaks, 
which determine the major part of the $k=1,\l=0$ moments. They are less prominent
for higher moments and consequently the amount of negative correlation increases with 
$\l=1,2,3$. This also explains the large and increasing positive correlations between the 
$k=1,\l\ge1$ moments, in which, with growing $\l$, the high energy tail is emphasised 
more than the low energy peaks. The total errors for the $(V+A)$ case are dominated by 
the uncertainties on the hadronic branching fractions.

\begin{table}[t]
\caption[.]{\label{tab:moments}
              Experimental ($D_{\tau, V/A}^{1\l}$) and theoretical ($D_{\tau, V/A}^{1\l\,({\rm theo})}$, 
              obtained after fit convergence, \cf Sec.~\ref{sec:fitResults})
              spectral moments of inclusive vector ($V$), axial-vector ($A$) and
              vector plus axial-vector $(V+A)$ hadronic $\tau$ decays.
              The errors $\Delta^{\mathrm{exp}} D_{\tau, V/A}^{1\l}$ summarise 
              statistical and systematic uncertainties.}
\begin{center}
\setlength{\tabcolsep}{0.0pc}
\begin{tabular*}{\columnwidth}{@{\extracolsep{\fill}}lcccc} 
\hline\noalign{\smallskip}
          &   $\l=0$   &   $\l=1$   &   $\l=2$   &   $\l=3$   \\ 
\noalign{\smallskip}\hline\noalign{\smallskip}
 $D_{\tau, V}^{1\l}$     &   0.71668  &   0.16930  &   0.05317  &   0.02254  \\ 
 $D_{\tau, V}^{1\l\,({\rm theo})}$     &   0.71568  &   0.16971  &   0.05327  &   0.02265  \\ 
 $\Delta^{\mathrm{exp}} D_{\tau, V}^{1\l}$      
                &   0.00250  &   0.00043  &   0.00054  &   0.00041  \\ 
\noalign{\smallskip}\hline\noalign{\smallskip}
 $D_{\tau, A}^{1\l}$     &   0.71011  &   0.14903  &   0.06586  &   0.03183  \\ 
 $D_{\tau, A}^{1\l\,({\rm theo})}$     &   0.71660  &   0.14571  &   0.06574  &   0.03130  \\ 
 $\Delta^{\mathrm{exp}} D_{\tau, A}^{1\l}$      
                &   0.00182  &   0.00063  &   0.00036  &   0.00025  \\ 
\noalign{\smallskip}\hline\noalign{\smallskip}
 $D_{\tau, V+A}^{1\l}$ &   0.71348  &   0.15942  &   0.05936  &   0.02707  \\ 
 $D_{\tau, V+A}^{1\l\,({\rm theo})}$     &   0.71668  &   0.15767  &   0.05926  &   0.02681  \\ 
 $\Delta^{\mathrm{exp}} D_{\tau, V+A}^{1\l}$  
                &   0.00159  &   0.00037  &   0.00033  &   0.00025  \\ 
\noalign{\smallskip}\hline\noalign{\smallskip}
  \end{tabular*}
  \end{center}
\vspace{0.3cm}
  \caption[.]{\label{tab:momcorr}
              Experimental correlations between the
              moments $D_{\tau,V/A/V+A}^{k\l}$. Correlations 
              between $R_{\tau, V+A}$, determined from the leptonic \Tau branching 
              fractions, and the corresponding moments are negligible.}
\setlength{\tabcolsep}{0.0pc}
\begin{tabular*}{0.32\columnwidth}{@{\extracolsep{\fill}}lrrrr} 
\hline\noalign{\smallskip}
& $D_{\tau,V}^{10}$ & $D_{\tau,V}^{11}$ 
& $D_{\tau,V}^{12}$ & $D_{\tau,V}^{13}$ \\ 
\noalign{\smallskip}\hline\noalign{\smallskip}
$R_{\tau,V}$      &$-0.287$ & $ 0.153$ & $ 0.274$ & $ 0.302$ \\
$D_{\tau,V}^{10}$ &  1     & $-0.821$ & $-0.981$ & $-0.993$ \\
$D_{\tau,V}^{11}$ & --     &  1      &  0.899   &    0.824 \\
$D_{\tau,V}^{12}$ & --     & --      &  1      &    0.988 \\
\noalign{\smallskip}\hline\noalign{\smallskip}
\end{tabular*}

\vspace{-3.065cm}\hspace{5.34cm}
\begin{tabular*}{0.32\columnwidth}{@{\extracolsep{\fill}}lrrrr} 
\noalign{\smallskip}\hline\noalign{\smallskip}
& $D_{\tau,A}^{10}$ & $D_{\tau,A}^{11}$ 
& $D_{\tau,A}^{12}$ & $D_{\tau,A}^{13}$ \\ 
\noalign{\smallskip}\hline\noalign{\smallskip}
$R_{\tau,A}$      &$-0.255$ & $ 0.013$ & $ 0.178$ & $ 0.272$ \\
$D_{\tau,A}^{10}$ &  1     & $-0.746$ & $-0.963$ & $-0.978$ \\
$D_{\tau,A}^{11}$ & --     &  1       &    0.866 &  0.646   \\
$D_{\tau,A}^{12}$ & --     & --      &    1      &  0.938   \\
\noalign{\smallskip}\hline\noalign{\smallskip}
\end{tabular*}

\vspace{-3.057cm}\hspace{10.8cm}
\begin{tabular*}{0.32\columnwidth}{@{\extracolsep{\fill}}lrrr} 
\noalign{\smallskip}\hline\noalign{\smallskip}
& $D_{\tau,V+A}^{11}$ 
& $D_{\tau,V+A}^{12}$ & $D_{\tau,V+A}^{13}$ \\
\noalign{\smallskip}\hline\noalign{\smallskip}
$D_{\tau,V+A}^{10}$ & $-0.722$ & $-0.974$ & $-0.987$ \\
$D_{\tau,V+A}^{11}$ &  1      &    0.801 &  0.662   \\
$D_{\tau,V+A}^{12}$ & --      &    1    &  0.975   \\
\noalign{\smallskip}\hline\noalign{\smallskip}
\vspace{0.3cm}
\end{tabular*}
\end{table}

\subsection{Fit Results}
\label{sec:fitResults}

\begin{figure}[t]
\centerline{\includegraphics[width=0.720\columnwidth]{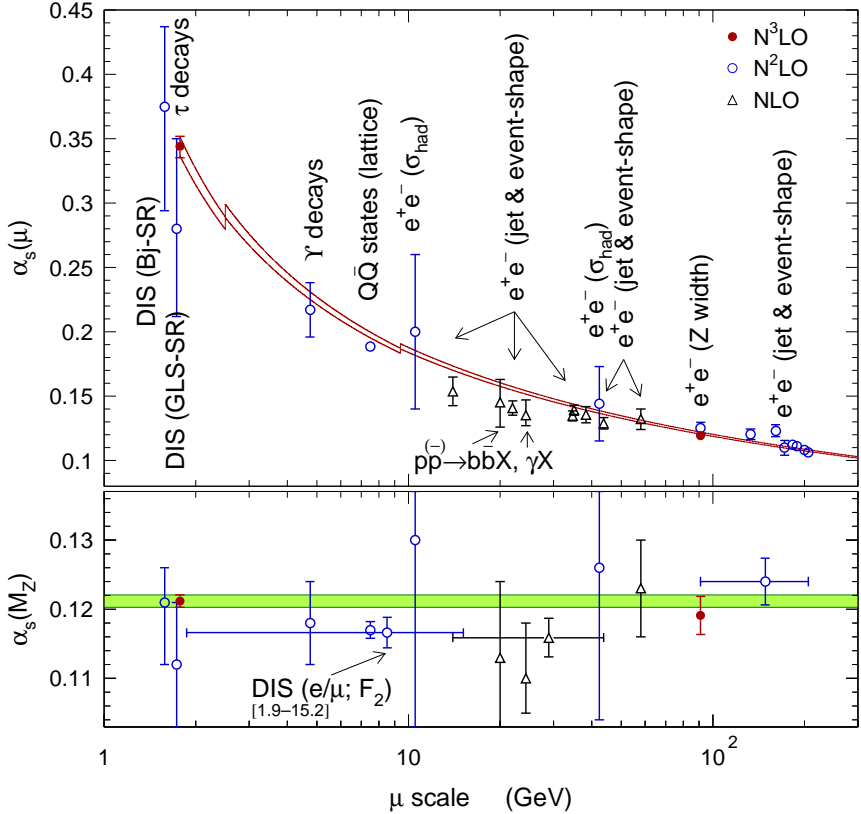}}
\vspace{-0.3cm}
\caption{Top: The evolution of \asTau to higher scales $\mu$ using 
         the four-loop RGE and the three-loop matching conditions applied at the
         heavy quark-pair thresholds (hence the discontinuities at $2\overline m_c$ 
         and $2\overline m_b$). The evolution is compared with independent 
         measurements (taken from the compilation~\cite{bethke2006}, and including 
         the recent measurements~\cite{eventShape,jade}) covering $\mu$ 
         scales that vary over more than two orders magnitude. Bottom: The corresponding 
         $\as$ values evolved to $\mZ$. The shaded band displays the $\tau$ decay 
         result within errors.}
\label{fig:evolution}
\end{figure}

Along the line of the previous analyses from ALEPH~\cite{aleph_as,aleph_asf,aleph_taubr,rmp}, 
CLEO~\cite{cleo_as}, and OPAL~\cite{opal_vasf}, we simultaneously determine \asTau, 
the gluon condensate, and the effective $D=6,8$ nonperturbative operators from a combined
fit to \Rtau and the spectral moments $D_{\tau,V/A}^{k\l}$ with $k=1$, $\l=0,1,2,3$,
taking into account the strong experimental and theoretical correlations between them.
\pskip
The fit minimises the $\chi^2$ of the differences between measured and
predicted quantities contracted with the inverse of the sum of the experimental
and theoretical covariance matrices. 
The theoretical uncertainties include separate variations of 
the unknown higher-order coefficient $K_5$, for which the value/error 
$K_5=K_4(K_4/K_3)\approx 378\pm378$ has been used, and of the renormalisation 
scale. The latter quantity has been varied within the range $\mtau^2 \pm2\,\gev^2$ 
(corresponding to $\xi=1\pm0.63$), and the maximum variations of the observables 
found within this interval are assigned as systematic uncertainties (\cf 
Sec.~\ref{sec:ptcomp}). To avoid double counting of errors the estimated $K_5$ term 
has been fixed when varying $\xi$. The corresponding systematic errors for 
\asTau are $0.0062$ ($K_5$) and $^{+0.0007}_{-0.0040}$ ($\xi$). The errors induced
by the uncertainties on $\Sew$ and $|V_{ud}|$ amount to $0.0007$ and $0.0005$, respectively.
With these inputs, the massless perturbative contribution $\delta^{(0)}$ is fully 
defined, and the parameter \asTau can be determined by the fit.
\begin{table*}[t]
  \caption[.]{\label{tab:alphas}
        Fit results for \asTau and the nonperturbative 
        contributions for vector, axial-vector and $V+A$ combined
        fits using the corresponding experimental hadronic widths and spectral moments 
        as input parameters, and using the CIPT prescription for the perturbative
        prediction. Where two errors are given the first is
        experimental and the second theoretical. The $\delta^{(2)}$ term 
        comes from theoretical input on the light quark masses varied within their 
        allowed ranges (see text). The quark condensates in the $\delta^{(4)}$ 
        term are obtained from PCAC, while the gluon condensate is 
        determined by the fit. The total nonperturbative contribution is 
        the sum $\delta_{\rm NP}=\delta^{(4)}+\delta^{(6)}+\delta^{(8)}$.}
\begin{center}
\setlength{\tabcolsep}{0.0pc}
\begin{tabular*}{\textwidth}{@{\extracolsep{\fill}}lccc} \hline 
\noalign{\smallskip}
  Parameter     &Vector ($V$) &Axial-Vector ($A$)&  $V\,+\,A$\\
\noalign{\smallskip}\hline\noalign{\smallskip}
 \asTau &  $0.3474\pm0.0074^{+0.0063}_{-0.0074}$  
        &  $0.3345\pm0.0078^{+0.0063}_{-0.0074}$   
        &  $0.3440\pm0.0046^{+0.0063}_{-0.0074}$   \\
\noalign{\smallskip}\hline\noalign{\smallskip}
  $\delta^{(0)}$  & $0.2093 \pm 0.0080$
                  & $0.1988 \pm 0.0087$
                  & $0.2066 \pm 0.0070$ \\
  $\delta^{(2)}$  & $(-3.2\pm3.0) \cdot 10^{-4}$
                  & $(-5.1\pm3.0) \cdot 10^{-4}$
                  & $(-4.3\pm2.0) \cdot 10^{-4}$ \\
\noalign{\smallskip}\hline\noalign{\smallskip}
 $\GG$ ($\gev^4$) &  $(-0.8\pm0.4) \cdot 10^{-2}$  
                  &  $(-2.2\pm0.4) \cdot 10^{-2}$   
                  &  $(-1.5\pm0.3) \cdot 10^{-2}$ \\
  $\delta^{(4)}$  & $(0.1\pm1.5) \cdot 10^{-4}$
                  & $(-5.9\pm0.1) \cdot 10^{-3}$
                  & $(-3.0\pm0.1) \cdot 10^{-3}$ \\
 $\delta^{(6)}$   &  $(2.68\pm0.20) \cdot 10^{-2}$  
                  &  $(-3.46\pm0.21) \cdot 10^{-2}$   
                  &  $(-3.7\pm1.7) \cdot 10^{-3}$   \\
 $\delta^{(8)}$   &  $(-8.0\pm0.5) \cdot 10^{-3}$  
                  &  $(9.5\pm0.5) \cdot 10^{-3}$   
                  &  $(8.1\pm3.6) \cdot 10^{-4}$   \\
  Total $\delta_{\rm NP}$ & $(1.89\pm0.25) \cdot 10^{-2}$  
                          & $(-3.11\pm0.16) \cdot 10^{-2}$
                          & $(-5.9\pm1.4) \cdot 10^{-3}$ \\
\noalign{\smallskip}\hline\noalign{\smallskip}
 $\chi^2/$DF  & 0.07   &  3.57   & 0.90 \\
\noalign{\smallskip}\hline
  \end{tabular*}
  \end{center}
\end{table*}
\pskip
Table~\ref{tab:alphas} summarises the results for the $V$, $A$ and $V+A$ combined 
fits using CIPT. The $\delta^{(2)}$ term is not determined by the fit, but is fixed 
from a theoretical input on the light quark masses varied within their errors~\cite{rmp}. 
The quark condensates in the $\delta^{(4)}$ term are obtained 
from partial conservation of the axial-vector current (PCAC), while the gluon
condensate is determined by the fit, as are the higher-dimensional operators
\Osix and \Oeight.
\pskip
The advantage of separating the vector and axial-vector channels and
comparing to the inclusive $V+A$ fit becomes obvious in the adjustment
of the leading nonperturbative contributions of $D=6$ and $D=8$, 
which have different signs for $V$ and $A$ and are thus suppressed in the 
inclusive sum. The total nonperturbative contribution, 
$\delta_{\rm NP}=\delta^{(4)}+\delta^{(6)}+\delta^{(8)}$, from the $V+A$ fit,
although non-zero, is significantly smaller than the corresponding values from 
the $V$ and $A$ fits, hence increasing the confidence in the \asTau determination 
from inclusive $V+A$ observables.
\pskip
There is a remarkable agreement within statistical errors between
the \asTau determinations using the vector and axial-vector data, with
$\asTauV-\asTauA=0.013\pm0.013$, where the error takes into account 
the anticorrelation in the experimental separation of the $V$ and $A$ modes. This 
result provides an important consistency check since the two corresponding 
spectral functions are experimentally almost independent, they manifest
a quite different resonant behaviour, and their fits yield relatively large
nonperturbative contributions compared to the $V+A$ case. 
Contrary to the vector case, the axial-vector fit has a poor $\chi^2$ value 
originating from a discrepancy between data and theory for the $\l=0,1$ 
normalised moments (\cf Table~\ref{tab:moments}). Although the origin of 
this discrepancy is unclear, it may indicate a shortcoming of the OPE in form
of noticeable inclusive duality violation in this channel. The observed systematic 
effect on the \asTau determination in this mode appears however to be within 
errors. From the fit to the $V+A$ $\tau$ spectral function, we obtain
\beqn
\label{astau-res}
 \asTau &=& 0.344 \pm0.005 \pm0.007~,
\eeqn
where the two errors are experimental and theoretical.
The values of the gluon condensate obtained in the $V$, $A$, and $V+A$ fits
are not very stable. Despite the apparent significance of the result for $V+A$, 
we prefer to enlarge the error taking into account the discrepancies between 
the $V/A$ results. We find for the combined value
$\GG = (-1.5 \pm 0.8)\cdot10^{-2}\,\gev^4$, which is at variance with the usual 
values quoted in the applications of SVZ sum rules. We note however that not much 
is known from theoretical grounds about the value of the gluon 
condensate~\cite{benekebraun}.
\pskip
The result~(\ref{astau-res}) can be compared with the recent 
determination~\cite{chetkuehn}, $\asTau=0.332\pm0.005\pm0.015$, also at \NNNLO, but 
using as experimental input only $R_{\tau,V+A}$, and not including 
the new information given in Sec.~\ref{sec:def}. Another major difference with our
analysis is that both perturbative procedures, FOPT and CIPT, are considered 
on equal footing, and their results are averaged. This leads to the lower 
value for \asTau and to an inflated theoretical error including half of the 
discrepancy between the two prescriptions.
\pskip
The evolution of the value (\ref{astau-res}) to $\mZ^2$, using Runge-Kutta 
integration of the four-loop $\beta$-function~\cite{rit}, and using
three-loop quark-flavour matching~\cite{chet12,wetzel1,wetzel2,pichsanta}, gives
\beqn
\label{eq:asres_mz}
   \asZtau &=& 0.1212 \pm 0.0005 \pm 0.0008 \pm 0.0005\:, \nonumber \\
                         &=& 0.1212 \pm 0.0011\:.
\eeqn
The first two errors in the upper line are propagated from the \asTau determination, 
and the last error summarises uncertainties in the evolution.\footnote
{The evolution error~\cite{rmp} receives contributions from the uncertainties 
   in the $c$-quark mass (0.00020, $\overline m_c$ varied by $\pm0.1\gev$) 
   and the $b$-quark mass (0.00005, $\overline m_b$ varied by $\pm0.1\gev$), the 
   matching scale (0.00023, $\mu$ varied between $0.7\,\overline m_q$ 
   and $3.0\,\overline m_q$), the three-loop truncation in the matching 
   expansion (0.00026) and the four-loop truncation in the RGE 
   equation (0.00031), where we used for the last two errors 
   the size of the highest known perturbative term as systematic 
   uncertainty. These errors have been added in quadrature.
}
All errors have been added in quadrature for the second line.
The result~(\ref{eq:asres_mz}) is a determination of the strong coupling
at the $Z$-mass scale with a precision of 0.9\%, unattained by any other 
\asZ measurement. The evolution path of \asTau is shown in the upper plot 
of Fig.~\ref{fig:evolution} (the two discontinuities are due to the chosen 
quark-flavour matching scale of $\mu=2\overline m_q$). The evolution is 
compared in this plot with other \as determinations compiled in~\cite{bethke2006} 
(we also included~\cite{jade}), and with new NNLO measurements based on hadronic 
event shapes from \ee annihilation covering the energy range between $91.2$ and 
$206\,\gev$~\cite{eventShape}. 
\pskip
The theoretically most robust precision determination of \as stems from the 
global fit to electroweak data at the $Z$-mass scale. As for \asTau, this 
determination benefits from the computation of the 
\NNNLO coefficient $K_4$ occurring in the radiator functions that predict the 
vector and axial-vector hadronic widths of the $Z$ (and also in the 
prediction of the total $W$ width). We use the newly developed {\em Gfitter} 
package~\cite{Gfitter} for the fit, and obtain
\beqn
\label{eq:alphasZ}
   \asZZ &=& 0.1191 \pm 0.0027 \pm 0.0001\:.
\eeqn
The value and first error represents the fit result, and the second error is due to the 
truncation of the perturbative series. It is estimated similarly to the \Tau case by 
adding a \fifth-order term proportional to $K_5$, estimated by $K_4(K_4/K_3)$, to the 
massless part, and a \fourth-order term (estimated accordingly), containing large 
logarithms $\ln(\overline m_t/\mZ)$, to the massive part. We also vary the
renormalisation scale of the massless contribution within the interval $\xi=1\pm0.63$, 
assuming the \fifth order coefficient to be known. The result~(\ref{eq:alphasZ}) agrees
with the finding of Ref.~\cite{chetkuehn}.
\pskip
The $\tau$-based result~(\ref{eq:asres_mz}) appears now twice more accurate than 
the determination from the $Z$ width. Yet the errors are very different in nature with 
a $\tau$ value dominated by theoretical uncertainties, whereas the determination 
at the $Z$ resonance, benefiting from the much larger energy scale 
and the correspondingly small uncertainties from the truncated 
perturbative expansion, is limited by the experimental precision 
of the electroweak observables. The consistency between the two results,
$\asZtau-\asZZ = 0.0021 \pm 0.0029$, provides the most powerful present test 
of the evolution of the strong interaction coupling as it is predicted by the 
nonabelian nature of QCD over a range of $s$ spanning more than three orders of 
magnitude. The \asZtau determination agrees with the average of the three currently 
most precise full NN(N)LO measurements (deep inelastic scattering~\cite{sy01,bethke2006}, 
ALEPH event shapes between 91 and 206 GeV~\cite{eventShape}, and global electroweak fit at \mZ), 
yielding an average of $0.1189\pm0.0015$ ($0.1204\pm0.0009$) when not including
(including) the \Tau result, which is justifiably assuming uncorrelated errors.
The \Tau-based result differs at the $2.5\sigma$ level from the value $0.1170\pm0.0012$ 
found in lattice QCD calculations with input from the mass splitting of the
$\Upsilon$ resonances~\cite{lattice}. The average of all five values reduces the 
discrepancy to $2.1\sigma$ ($\chi^2$ probability of 0.04).

\section{Conclusions}\label{sec:conclusions}

We have revisited the determination of \asTau from
the ALEPH $\tau$ spectral functions using recently available results. On the
experimental side, new BABAR measurements of the $e^+e^-$ annihilation cross 
section into $K\Kb\pi$ using the radiative return method now
permit, through CVC, a much more accurate determination of the
vector/axial-vector fractions in the corresponding $\tau$ decays.
Also, better results are available on $\tau$ decays into strange final states
from BABAR and Belle. On the theory side, the first unknown term in the
perturbative expansion of the Adler function, the \fourth-order term $K_4$,
was recently calculated, opening the possibility to further push the
accuracy of the theoretical analysis of the hadronic $\tau$ decay rate.
 \pskip
Motivated by these improvements we have reexamined the theoretical 
framework of the analysis. In particular the convergence properties of the
perturbative expansions for the \Tau and $Z$ hadronic widths have been studied, 
and the ambiguity between the fixed-order (FOPT) and contour-improved
(CIPT) approaches for summing up the series has been discussed. The study
confirms our earlier findings (at \third order) that CIPT is the more reliable
treatment. Furthermore we have identified specific consistency problems
of FOPT, which do not exist in CIPT. Possible violations of quark-hadron 
duality at the \Tau mass scale have been considered using specific models, 
and their effect has been found to be well within our quoted overall theoretical 
uncertainty (however, due to the coarseness of the models, we do not introduce
additional theoretical errors).
\pskip 
We perform a combined fit of the \Tau hadronic width and hadronic spectral moments
resulting in the value $\asTau=0.344\pm0.005_{\rm exp}\pm0.007_{\rm theo}$,
consistent with the previous value obtained for three known orders, and with 
a 20\% reduced theoretical uncertainty. This somewhat moderate improvement
is the result of the relatively large value $K_4\sim49$, suggesting a slowly
converging perturbative series and giving rise to relatively large truncation
uncertainties. Nevertheless, the result confirms the excellent accuracy that can
be obtained from the analysis of $\tau$ decays, albeit indicating that
this method may approach its ultimate accuracy.
\pskip 
The evolved \Tau result at the \mZ scale,
$\asZ=0.1212 \pm 0.0005_{\rm exp}\pm 0.0008_{\rm theo}\pm 0.0005_{\rm evol}$,
is the most accurate determination available. It agrees with the corresponding
value directly obtained from $Z$ decays, which we have reevaluated. Both
determinations are so far the only results obtained at \NNNLO order. They confirm
the running of $\as$ between 1.8 and 91\,\gev as predicted by QCD with an
unprecedented precision of 2.4\%.
 
%
%
\begin{details}
We are indebted to Martin G\"obel and the Gfitter group for implementing the new
\NNNLO term into the global electroweak fit, and AH acknowledges the fruitful 
collaboration. We thank Oscar Cat\`{a}, Maarten Golterman and Santi Peris for 
letting us preview their analysis on duality violation in hadronic \Tau decays 
(which arrived after finalising this paper), and for helpful discussions.
Many thanks to Matthias Jamin for pointing out a mistake in the total 
nonperturbative contributions previously quoted in Table~\ref{tab:alphas} (corrected
in the present version). This work was supported in part by the EU Contract No. 
MRTN-CT-2006-035482, \lq\lq FLAVIAnet''.
\end{details}

\end{document}